\documentclass[pdflatex,sn-mathphys-num]{sn-jnl}

\usepackage[utf8]{inputenc}
\usepackage[T1]{fontenc}
\usepackage{natbib}
\usepackage{amsmath}
\usepackage{color}
\usepackage{graphicx}
\usepackage{csquotes}
\usepackage{cleveref}
\usepackage{booktabs}
\usepackage{tabularx}

\title{Anticipatory Human Oversight of Agentic AI:\\A Philosophical Account}

\author*[1,2,3]{\fnm{Kevin} \sur{Baum}}\email{kevin.baum@dfki.de}

\author[2]{\fnm{Maximilian} \sur{Kiener}}

\author[4]{\fnm{Markus} \sur{Langer}}

\author[3]{\fnm{Johann} \sur{Laux}}

\affil*[1]{\orgname{German Research Center for Artificial Intelligence (DFKI)}, \orgaddress{\city{Saarbr\"ucken}, \country{Germany}}}

\affil[2]{\orgdiv{Institute for Ethics in Technology}, \orgname{Hamburg University of Technology (TUHH)}, \orgaddress{\city{Hamburg}, \country{Germany}}}

\affil[3]{\orgdiv{Oxford Internet Institute}, \orgname{University of Oxford}, \orgaddress{\city{Oxford}, \country{United Kingdom}}}

\affil[4]{\orgdiv{Department of Psychology}, \orgname{University of Freiburg}, \orgaddress{\city{Freiburg}, \country{Germany}}}

\begin{document}
\maketitle
\begin{abstract}
\noindent
Human oversight is widely held to mitigate the 
risks of 
AI systems. Even for systems that produce discrete outputs at identifiable decision points, the realisation of human oversight as a reactive measure is empirically fragile, yet increasingly well understood. However, for agentic AI---systems that plan, decompose goals, and execute multi-step actions over extended horizons---reactive oversight reaches its structural limits: intervention on individual actions defeats the autonomy that motivates the deployment, while intervention on aggregate patterns is too coarse for harms whose cumulative consequences only become legible after the fact.

This paper argues that reactive oversight must be complemented by an \emph{anticipatory} mode: oversight exercised before the agent acts, by specifying the \emph{normative agenda} that structures the space of permissible action and refining it iteratively through specification, runtime, and inspection. The two are complements---the agenda's escalation conditions specify when reactive intervention is invoked.

Drawing on Meaningful Human Control, we read anticipatory oversight as the operationalisation of distal-reason tracking. In addition, we argue that the proposed framework yields a specific responsibility architecture by design: occupying the anticipatory mode is the discharge of a role-grounded prospective obligation, and backward-looking responsibility takes the form of strict moral answerability---rationalistic, relational, and holding regardless of fault, in virtue of the principal's prior opportunity for precaution. We develop bridging failure modes, 
address objections including moral luck and the illusion of control, and close with regulatory, architectural, and empirical implications

\end{abstract}

\section{Introduction}
\label{sec:intro}

AI agents are AI systems built to operate with a degree of autonomy: rather than only \textit{answering}, preparing information, or asking permission at each step, they pursue goals by \textit{acting}.
While artificial agents are not a new topic, the capabilities of recent systems---planning multi-step procedures, decomposing high-level goals into sub-goals, coordinating with other systems, using tools, and selecting actions in open-ended environments with limited human supervision---have a new quality \citep{botti2025agentic,gabriel2024ethics, chan2024visibility, kasirzadeh2025characterizing,Dung2025-DUNUAA}. The shift from systems-that-answer to systems-that-act fulfils a longstanding aspiration of AI research and raises new challenges for AI ethics, in particular for our understanding of responsibility and effective human control. This article uses the established concept of human oversight to offer a new stance on these challenges, by asking what effective and meaningful human oversight looks like when the systems being overseen no longer present discrete, identifiable decision points at which a human can intervene, but generate streams of micro-actions whose individual stakes may be low, whose cumulative consequences can still be severe, and whose continuous review would defeat the very autonomy that motivates the deployment.

The question matters for AI governance and regulation. Article~14 of the EU AI Act introduces human oversight as a central safeguard for high-risk AI systems \cite{euaiact2024}, and most major AI governance frameworks treat human oversight as indispensable \citep{fink2025oversight, enarsson2023oversight, sterz2024quest}. The conceptual models these regimes rely on---typically considered to be some form of \emph{human-in-the-loop} (HITL) or \emph{human-on-the-loop} (HOTL)---%
were developed for systems whose decision cycles, even where continuous, present identifiable points of human intervention
\citep{baum2026causal,laux2023institutionalised}.
A substantial empirical literature documents that even for systems of this kind, the effectiveness of human oversight is not trivial to achieve \citep{green2022flaws,parasuraman2010complacency, sterz2024quest, langer2025signal, elish2019moral}. For agentic AI, the difficulty is more than empirical: the reactive operation these models presuppose---intervention on the agent's actions, in real time or after the fact, whether on individual outputs or on aggregated patterns---reaches its structural limits. Continuous intervention on individual actions degrades the autonomy that motivates the deployment of agentic systems; intervention on aggregated patterns is too coarse to prevent harms whose pattern only becomes legible after the fact \citep{passi2025agentic, mitchell2025fully, shapira2026agents}.

In reaction, recent work has begun to turn to a third mode of human involvement, distinct from real-time monitoring and from periodic audit. 
In the five-way taxonomy of \citet{singh2025architecting}, a new \emph{Human-over-the-Loop} category names, as \citet{lazaros2026human} characterise it, \enquote{a configuration where humans are in a position of power with respect to system goals and constraints}.
Similarly, in the causal taxonomy of \citet{baum2026causal}, a new variant of HOTL---\emph{anticipatory HOTL}---names a structurally distinct mode of corrective human involvement: a human \enquote{exercises corrective influence prospectively by establishing normative constraints, authorization boundaries, or alignment parameters before the system operates}. The mode operates on the conditions under which the agent acts rather than on its individual outputs.  
A recent consensus paper from the 2025 Dagstuhl seminar on human oversight of AI reaches a similar diagnosis: for highly autonomous systems where real-time oversight \enquote{may be infeasible}, it identifies \enquote{containment architectures that enable anticipatory oversight by providing systems with guardrails before deployment} as a research frontier \citep{gaube2026keeping}.  

This paper develops a philosophical account of what occupying the anticipatory mode substantively requires for agentic AI. The category is already marked out by the work surveyed above; our contribution is to give it conceptual and normative clarity and shed light on its implications for responsibility.

Concretely, the proposal we develop runs as follows. Instead of monitoring the agent's individual actions as they occur or in aggregated form and intervening in reaction to an unsafe signal \citep{langer2025signal}, the principal specifies in advance---before the agent's engagement---a \emph{normative agenda}: a structured set of rules, constraints, values, and escalation conditions under which the agent operates. The agent then plans, decomposes goals, and acts within this structured normative space. After the engagement (or after an escalation condition is triggered), the principal inspects what happened, reviews logs and justifications the system produced for its choices, and refines the agenda for the next iteration in light of what inspection surfaces. Thus, the principal still exercises causal influence on the system's behaviour, but by shaping the conditions under which the agent acts---not through real-time intervention on individual actions, and not through retrospective reaction to aggregated outputs alone.
Anticipatory oversight extends rather than displaces reactive oversight: the agenda's escalation conditions---tripwires, uncertainty thresholds, triggered hand-offs, and the like---specify when reactive intervention during the engagement is to be invoked, on terms the agenda has fixed in advance. This is still human oversight in the substantive sense---a runtime risk-mitigation practice, arguably in line with Article~14 of the EU AI Act as we argue in \Cref{sec:implications}---yet the locus of the principal's causal contribution has shifted.

Three substantive claims follow from this proposal, which the rest of the paper develops. First, anticipatory oversight is best understood, in the Meaningful Human Control (MHC) tradition, as the operationalisation for agentic AI of the tracking of \emph{distal} reasons---the plans, values, and role-typical commitments that are temporally and semantically remote from any single action, and that proximal tracking leaves systematically unaddressed once systems act over time \citep{santoni2018meaningful, mecacci2020meaningful}; normative agenda-setting is the mechanism through which such tracking becomes tractable \citep{baum2026principles}.  
Second, the regime makes a specific architecture of responsibility available: occupying the anticipatory mode is itself the discharge of a role-grounded prospective obligation \citep{hardimon1994, vandepoel2011relation, duff2007}, and the corresponding backward-looking responsibility takes the form of \emph{strict moral answerability} \citep{kiener2024strict, kiener2025abundance}---rationalistic, relational, and holding in virtue of the principal's prior opportunity for precaution rather than fault.
Third, the framework supports a taxonomy of failure modes---bridging failures of specification, translation, compliance, and inspection---each pointing to a distinct locus of responsibility when something does go wrong. The aim, throughout, is to do for the anticipatory mode what the literature on effective human oversight has done for reactive modes \citep{sterz2024quest, langer2025signal, gaube2026keeping}:  
identify the conditions under which the position is genuinely occupied and the role genuinely discharged, rather than merely formally assigned.

A running example helps fix ideas. \emph{The Secretary Agent.} A principal deploys a personal AI agent that manages her professional correspondence: \enquote*{triaging} messages, drafting routine replies, scheduling meetings, declining low-priority requests, escalating items that require her attention. The agent operates continuously, handling dozens of items per hour. The principal is a research-group lead with varied correspondents---collaborators, students, journalists, administrative staff---and varied confidentiality requirements. Real-time review of individual actions would be substantive only by collapsing the deployment into a sequence of approved single decisions, defeating the agent's autonomy and dissolving the very efficiency gain that motivated delegating to an agent in the first place; periodic review of aggregate patterns preserves that autonomy but is too coarse to capture the principal's normative commitments. Neither, then, gives her substantive oversight of an \emph{agentic} system. What structure of oversight would let the agent act with autonomy while keeping her normative commitments in force? The example returns throughout the paper.

For agentic AI the structural problem is to provide substantive human control without the monitoring that oversight has classically meant. The rest of the paper develops what occupying the anticipatory mode requires. \Cref{sec:locating} lays the foundations, drawing on the supervisory-control tradition, recent taxonomies of human runtime involvement, and the MHC framework. \Cref{sec:agenda} specifies the normative agenda, its iterative refinement, and the architectural desiderata for the alignment and containment mechanisms that operationalise it. 
\Cref{sec:responsibility} sets out the responsibility architecture, with strict moral answerability as the form responsibility takes, and develops the failure-mode taxonomy as the dual of its responsibility constellations. \Cref{sec:limits} addresses objections and limits, including the redistribution of moral luck and the risk of an illusion of control.
\Cref{sec:implications} draws out implications for the EU AI Act, layered governance, and architectural and empirical research.
\Cref{sec:conclusion} concludes.

\section{Foundations of Anticipatory Oversight}
\label{sec:locating}

This section lays the foundations on which \Cref{sec:agenda} builds. There are two. The first is \emph{architectural}: occupying the anticipatory mode has to inherit from the traditions in which humans have classically maintained substantive control over delegated activity (\Cref{subsec:traditions}), via the runtime-internal structure that agentic operation actually presents (\Cref{subsec:runtime}). The second is \emph{philosophical}: what substantive occupancy of the mode requires is intelligible against the Meaningful Human Control (MHC) framework, whose two conditions---tracking and tracing---are reconstructed for the agentic case (\Cref{subsec:mhc}). The closing subsection (\Cref{subsec:residual}) names what these foundations leave to be developed in \Cref{sec:agenda}.

\subsection{Oversight as Practice: Three Traditions}
\label{subsec:traditions}

We begin with the architectural foundation. 
The structural problem set out in \Cref{sec:intro} is not entirely new. Structurally adjacent problems---how a human maintains substantive control over a process she no longer acts on moment-to-moment---have been addressed in several traditions whose joint contribution to the anticipatory mode this section synthesises. Three are central. From human factors, Sheridan's supervisory control theory characterises the role of the human supervisor in systems with high levels of automation \cite{sheridan1978human,sheridan1992telerobotics,sheridan2021supervisory}. From management theory, the literature on delegation works out how a principal maintains substantive control over delegated activity \cite{drucker1954practice}. From economics, principal-agent theory formalises the structure of delegation under information asymmetry \cite{jensen1976,eisenhardt1989agency}. Each tradition captures part of what the anticipatory mode requires; none on its own delivers the philosophical substance the agentic case demands; and one---principal-agent theory---imports assumptions we will need to fence rather than inherit. 
The rest of the subsection develops a synthesis of all three, plus a further move the traditions can suggest but cannot themselves deliver.

Sheridan's framework characterises supervisory control as the regime that emerges when humans cease to operate a process directly and instead supervise an automated system performing it. The shift is from operator to supervisor: as automation grows, the human's role moves away from moment-to-moment control toward functions that bracket the automated execution. Sheridan decomposes the supervisor's role into five sequential functions. The supervisor \emph{plans} offline what task is to be done and how; \emph{teaches} the computer what was planned, in the agentic case by configuring and prompting the agent; \emph{monitors} the automatic action as it unfolds, to detect deviations or failures; \emph{intervenes} when intervention is required; and \emph{learns} from each engagement what to do better next. 
Automation itself reduces what monitoring can feasibly deliver: the more automated the process, the less the supervisor can monitor without defeating the very autonomy that motivated automation in the first place.
And monitoring is in any case the function most difficult to perform well over time:
attention drifts, anomalies become familiar, vigilance erodes \citep{bainbridge1983ironies,parasuraman2010complacency}. The substance of effective supervision thus shifts upstream and downstream---to planning and teaching before the engagement, and to learning after. This is the structural insight the anticipatory mode for agentic AI inherits.

Management theory and economics work the same delegation pattern in different vocabularies. Drucker's canonical move in modern management is that a manager who delegates a complex task does not intervene during the subordinate's work but specifies goals and constraints beforehand and assesses performance afterwards \citep{drucker1954practice}. The model takes for granted that the subordinate is a competent agent whose understanding of the principal's expectations can be developed through professional socialisation, training, and conversation. 

Principal-agent theory, developed in economics, formalises the same delegation structure under conditions of information asymmetry, with the principal's central tools being the specification of authorised actions, reporting mechanisms, and constraints under which deviation can be detected and corrected \citep{jensen1976,eisenhardt1989agency}. The formal delegation vocabulary transfers cleanly to the agentic case; its substantive assumption---that the agent has interests potentially divergent from the principal's that must be aligned through incentives---does not. For agentic AI we substitute misalignment and behavioural drift: the agent is not pursuing different interests from the principal but may have inherited or developed dispositions that diverge from the agenda she has set.

Each of the three traditions was developed for a context that translates only imperfectly to agentic AI. 
Sheridan's framework was built for cyberphysical systems with clear ground truth; 
the management tradition for human subordinates whose understanding can be checked in conversation; principal-agent theory for utility-maximising agents in contractual relationships. The most important divergence is the second. Human delegation, on the management-theory model, works through \emph{checkable shared understanding}: distal reasons---the plans, values, and role-typical commitments the principal brings to the engagement---travel through professional socialisation and tacit organisational knowledge; the manager can recalibrate in conversation when the subordinate's reading drifts; the subordinate can be asked \enquote{do you see why this matters} and can answer. None of these channels is reliably available in the agentic case: an AI agent has no professional socialisation through which distal reasons accumulate, no tacit organisational knowledge to inherit, and no guarantee that conversation will recalibrate its underlying dispositions. The distal reasons that human delegation carries through socialisation and conversation therefore have to be made structurally explicit \emph{ex ante} for an agentic AI, in a form the agent can track and the principal can later inspect. This is the move the traditions can suggest but cannot themselves deliver, and it is what the rest of the paper develops.

\subsection{The Runtime of Agentic Operation}
\label{subsec:runtime}

While other forms of oversight matter across the broader lifecycle of an AI system---design reviews, audits, post-market monitoring, organisational governance \citep{gaube2026keeping}---\emph{human oversight} in the sense at issue here is a runtime risk-mitigation practice \citep{enarsson2023oversight, fink2025oversight, sterz2024quest, langer2025signal}, concerned with the period in which the system is operationally deployed and acts in the world. The runtime of an agentic AI system, however, has its own internal structure that mirrors the system's lifecycle in miniature---a supervisory-control loop in its own right, with its own specification, action, and inspection phases (see also \Cref{fig:spiral} below). A specific instance of agentic operation begins with a configuration- and prompt-laden \emph{specification phase}, in which the principal formulates the agent's task and the constraints under which it is to be carried out; the agent then plans, decomposes goals, and calls tools---it \emph{acts}; and the engagement closes either with a delivered result or with a failure of some kind, on the basis of which the principal can \emph{inspect}, learn, and refine the next iteration.

Mapped onto Sheridan's five-function characterisation of supervisory control, the three phases of the agentic runtime distribute as follows. The specification phase is where the principal \emph{plans} and \emph{teaches}---formulating the engagement's task and the constraints under which the agent is to operate. The action phase is where the agent's autonomy lives, and where reactive monitoring and intervention would, on Sheridan's framework, sit. The inspection phase is where the principal \emph{learns} from the engagement and refines the agenda for the next iteration. 
The structural problem set out in \Cref{sec:intro} lands precisely on the monitoring function during the action phase: the very feature that makes monitoring effective elsewhere defeats the autonomy that motivates the agentic deployment in the first place, while the looser alternative leaves \emph{trajectory-level} failures unaddressed---cumulative failures whose shape emerges only across many of the agent's actions taken together.
The substance of effective supervision therefore redistributes upstream and downstream, to plan-teach in specification and to learn in inspection. The anticipatory mode lives in this redistribution.

This three-phase structure of the runtime of an agentic AI system is the territory the recent taxonomies surveyed in \Cref{sec:intro} have converged on. Singh and Szajnfarber's Human-\textit{over}-the-Loop \citep{singh2025architecting}, Baum and Laux's anticipatory HOTL \citep{baum2026causal}, and the Dagstuhl consensus on containment architectures \citep{gaube2026keeping} all identify the same conceptual object: a mode of human involvement that exercises control upstream and downstream of the agent's action rather than during it. 
What none of them develops is the substance of what occupying that mode philosophically requires---the question of which reasons the agent's behaviour must track, on whose authority, and through what apparatus. \Cref{subsec:mhc} takes up that question via the Meaningful Human Control framework.

\subsection{Meaningful Human Control and the Reasons Foundation}
\label{subsec:mhc}

With the architectural foundation in place, we turn to the philosophical one. The Meaningful Human Control framework specifies two conditions on an automated system \citep{santoni2018meaningful}. The first is \emph{tracking}: the system's behaviour systematically covaries with the relevant reasons of the relevant human agents. The second is \emph{tracing}: there is at least one human in the design or operation chain who understands the system's capabilities and is appropriately positioned for moral responsibility for what it does. Both conditions are framework-level requirements; both are demanding for agentic AI in ways the framework's initial formulations did not anticipate. Recent work has made the two-condition structure explicit and worked through the difficulties for contemporary systems \citep{kozlovski2025reasons}. The present subsection lays the framework's foundations, which the rest of the paper builds on: \Cref{sec:agenda} reconstructs tracking for agentic AI, and \Cref{sec:responsibility} reconstructs tracing.

The notion of tracking is technical. A system tracks a reason just in case its behaviour would be different if the reason were different. Santoni de Sio and van den Hoven formalise this counterfactual relation via Nozick-style subjunctive conditionals and read it as a form of reasons-responsiveness.\footnote{For the subjunctive-conditional reading, see \cite{Nozick1981}; for the reasons-responsiveness tradition, \cite{fischer1998}.} The substantive question this raises---which the framework leaves open---is \emph{which} reasons of \emph{which} agents a system ought to track. The substance of meaningful control turns on the answer.

\citet{mecacci2020meaningful} answer this question by distinguishing reasons by their temporal and semantic distance from the action they bear on.\footnote{The distinction adopts the proximal/distal labels with which \citet{mele1992springs} renders Bratman's dichotomy of intentions, against the background of the Anscombe-Davidson-Bratman tradition of theorising action in terms of reasons, intentions, and plans \citep{anscombe1957intention, davidson1980essays, bratman1987intention}.} \emph{Proximal} reasons are temporally close to and semantically narrow with respect to the action---the intention to brake now, to send this draft, to schedule this meeting. \emph{Distal} reasons are temporally more removed and semantically broader: plans in Bratman's sense, role-typical commitments, values, and institutional norms---the plan to arrive home safely, the commitment to communicate with collaborators in a way that maintains the research group's standing, the norm of confidentiality of unpublished work. 
The distinction is orthogonal to that between motivating and normative reasons: both proximal and distal reasons can be either.\footnote{Roughly speaking, \textit{motivating} reasons are the considerations an agent takes herself to be acting on; \textit{normative} reasons are considerations that count in favour of (or against) an action regardless of whether the agent so acts \citep{dancy2000practical, scanlon1998, raz1990practical}.}
What distinguishes them is the scope and timeframe of the action with respect to which they bear. Mecacci illustrates this with the dual-mode vehicle: a self-driving car can be highly responsive to proximal driving reasons---the reason to slow now, to merge here, to yield to this pedestrian---and still fail to track the distal reasons that organise the trip, such as the driver's plan to arrive home safely rather than quickly. The asymmetry is the substantive point: proximal tracking can be impeccable while distal tracking fails.

\citet{veluwenkamp2022reasons} sharpens the framework on the normative side. The reasons MHC tracking must track, he argues, are normative rather than motivating: a system responsive only to the considerations its principal happens to be moved by has not been brought under meaningful control in the substantive sense the framework requires. The point matters because most of the distal reasons the agenda encodes are normative: the institutional confidentiality norm, the role-typical mentorship commitment, the value-bearing categories the principal's role makes binding for her are not just considerations she finds motivating but considerations that hold of her independently of motivation. The normative-reasons reading also raises a worry \citet{veluwenkamp2022reasons} himself develops. If normative reasons can be agent-neutral---as on consequentialist accounts which, slightly simplified, treat whatever makes the world better as a reason for anyone to act---then tracking such reasons cannot fix \emph{whose} reasons the system tracks, and therefore cannot do the responsibility-attribution work that gives the framework its point. The worry is real for any reading of MHC that grounds responsibility on tracking alone. We take it up in \Cref{subsec:roles}: the distal reasons the agenda encodes are role-indexed for the role-occupant whose delegation it governs, and so agent-relative by construction.

The framework's task takes its place within an established responsibility-gap discourse.\footnote{\citet{santoni2021four} articulates four interconnected responsibility gaps with which any account of meaningful control must reckon; the responsibility architecture the present framework makes available is developed against this structure in \Cref{sec:responsibility}.} On the reading we adopt, however, the substantive task is better described as closing a \emph{control-gap}: not locating someone to blame for harm but building systems whose level of risk is morally acceptable \citep{hindriks2023control}. The anticipatory mode is exactly what closing such a gap looks like for agentic AI.\footnote{A diagnosis recently echoed from the engineering side by \citet{li2026controllability}, who argue runtime controllability (interruptibility, override authority, persistence, auditability) must be treated as a first-class design objective alongside alignment.}

\subsection{What Remains to Show}
\label{subsec:residual}

The two foundations are now in place. The architectural inheritance from the supervisory-control and delegation traditions locates the substance of oversight in the specification and inspection phases of the agentic runtime, away from the structurally compromised monitoring loop. The Meaningful Human Control framework supplies the philosophical substance: tracking, in the technical sense, against the distal reasons the principal's role makes normative for her---with the agent-neutrality worry resolved by the role-indexed structure of those reasons. What remains is to develop the substantive operationalisation: how distal reasons are structured into a form the agent can track; what work the principal's roles do in that structuring; how iterative refinement makes the regime stable; and what architectural demands the alignment and containment mechanisms must meet. \Cref{sec:agenda} develops each of these in turn.

\section{Operationalising the Anticipatory Mode}
\label{sec:agenda}

\Cref{sec:locating} laid the foundations: the architectural inheritance from oversight and delegation traditions, and the philosophical apparatus of the Meaningful Human Control framework. This section develops what occupying the anticipatory mode substantively requires. \Cref{subsec:normative-agenda} introduces the \emph{normative agenda} as the operationalisation of distal-reason tracking and characterises its content and the mechanism families through which it binds the agent. \Cref{subsec:roles} develops the role-grounded structure that makes the agenda's distal reasons agent-relative and answers the agent-neutrality worry raised in \Cref{subsec:mhc}. \Cref{subsec:reasons} treats the agenda as a structure of normative reasons whose interaction is defeasible and non-monotonic. \Cref{subsec:spiral} sets out the design--runtime--inspection \emph{spiral} as the iterative form anticipatory oversight takes. \Cref{subsec:desiderata} states the three architectural desiderata the agenda's substrate must meet.

\subsection{The Normative Agenda}
\label{subsec:normative-agenda}

For agentic AI the asymmetry Mecacci's dual-mode vehicle illustrates is sharper still. An agent that plans over extended horizons, decomposes high-level goals into sub-goals, and selects actions in open-ended environments generates a stream of individually small decisions whose cumulative shape is the locus of the morally significant behaviour. Each individual action---a message triaged, a reply drafted, a meeting declined, an API called---can be entirely defensible in its local context while the trajectory through which the agent moves across many such actions departs systematically from what the principal would endorse.

Consider the secretary agent. Over a working week, each individual meeting it declines on the principal's behalf is locally well-justified: the slot is already taken, the requester is unfamiliar, the topic is peripheral. Across the week, however, the pattern of declines falls disproportionately on student-supervision requests, because those tend to arrive on short notice and to compete with longer-standing journalist and collaborator commitments---a trajectory the principal, whose role-typical commitment to mentorship is a distal reason of high standing, would on reflection repudiate, but which no individual decline brought into view. Or: each time the agent forwards an unpublished manuscript as context for an ongoing exchange, there is a local reason to do so (the correspondent asked, the manuscript bears on the question, the recipient is in principle a peer), and yet the cumulative effect across half a dozen such forwards is a circulation of confidential work at a scale the principal's confidentiality norm rules out at the trajectory level, even where it does not rule out the individual exchange.

Reactive oversight, whether exercised in real time or after the fact, is at home in this proximal register---the operational zone of individual actions and their locally relevant reasons. What it cannot do, structurally, is set the distal reasons against which the trajectory as a whole is to be assessed: those have to be in place before the agent starts acting, in a form the agent can track. Anticipatory oversight is, on this reading, the operationalisation of distal-reason tracking for agentic AI. The principal's distal reasons---her plans, her role-typical commitments, the institutional norms her deployment context puts in force---are made structurally available to the agent as the conditions under which it acts, and are refined iteratively as inspection of trajectories surfaces where the present articulation is too coarse, too permissive, or too brittle.

What does operationalising distal-reason tracking plausibly look like? The distal reasons in question---the principal's plans, role-typical commitments, values, and the institutional norms her deployment puts in force---are not a finite list of rules or a closed set of forbidden states; they are an open, partially articulable texture against which the propriety of trajectories is assessed. A mechanism demanding the principal exhaustively enumerate them in advance would inherit precisely the problem it was meant to solve. What the anticipatory mode requires is not complete specification but \emph{articulated} specification: a structured representation of those distal reasons the principal can and does make explicit, designed to be revisable as inspection of trajectories surfaces what she has not yet articulated or has articulated too coarsely. We call this the \emph{normative agenda} of the deployment, following the account of principal-agent delegation developed in \citet{baum2026principles}: the explicit, documented, and supervisable specification of the normative conditions under which the principal maintains a delegation---the principal's structured articulation of distal reasons for this deployment, in this context, at this time, in a form the agent can track and the principal can later inspect, justify, revise, and extend.  

The work the agenda does in a deployment is two-sided. As \emph{agent-facing} structure, it constitutes the conditions under which the agent acts: whatever mechanism mediates between agenda and agent counts as the realisation of anticipatory oversight only insofar as it operationalises the agenda in a form the agent can track in the technical sense---the agent's behaviour covaries with the agenda when the agenda changes. As \emph{principal-facing} structure, the agenda is what the principal herself is committed to, what her inspection of the agent's trajectories is calibrated against, and what other parties may, where appropriate, hold her to account for. Anticipatory oversight thus binds the principal to an articulated commitment as much as it binds the agent to act within that commitment's terms: an agenda is not just \emph{set}, it \emph{holds}.

Normative agendas are operationalised through three complementary mechanism families. \emph{Alignment} shapes the agent's responsiveness to reasons---the internal dispositional structure that makes the agent track the agenda as a matter of its own behaviour \citep{gabriel2020values, 10.1007/978-3-032-01377-4_8}.  
\emph{Internal containment} is structural-architectural: features within the agent's decision pipeline that constrain what it can do regardless of its disposition---capability gates on tool-use, sub-agent sandboxing, or architectural separations of normative reasoning from instrumental optimisation, exemplified by \citet{jahn2026grace}'s neuro-symbolic architecture.  
\emph{External containment} is institutional and infrastructural \citep{gaube2026keeping, li2026controllability}: operating-system sandboxes, network-level capability limits, runtime monitors that fire tripwires, uncertainty-threshold detectors that trigger escalation, hand-offs that bring a human back into the loop.  
The three families are complementary, and the agenda's binding force on agent behaviour during the action phase is their conjunction. The agenda is the input to these mechanisms, not a substitute for one: an agenda articulated but to which no available mechanism can render the agent responsive is articulated intent without operative anticipatory oversight.

An agenda articulated for a real deployment typically contains four kinds of content. The \emph{purposes} of the delegation specify what the agent is to pursue, at what granularity, and with what priority where purposes conflict. \emph{Authorisation boundaries} specify what the agent may and may not do in service of those purposes---which actions, instruments, data, or counterparties are in or out of scope, and under what conditions. The \emph{considerations} the agenda identifies are the distal reasons made explicit at the granularity the principal can articulate them at this stage---the values, role-typical commitments, and institutional norms against which candidate actions are to be assessed when local rules underdetermine the case. \emph{Escalation triggers} specify the conditions under which the agent is to defer to a human or hand control back: uncertainty thresholds, novelty conditions, stake-related flags, and trajectory-level drift indicators. They are the explicit points at which anticipatory oversight calls reactive oversight back into the loop on terms the agenda has fixed in advance---the primary operational interface between the two complementary modes. On the substrate side, they are implemented through external containment as set out in the previous paragraph. None of the four components is sharply disjoint from the others, and a non-trivial agenda will contain content of all four kinds in interlocking form.\footnote{These four content kinds structure the agenda. The failure modes of \Cref{sec:failures}, by contrast, are keyed to the phases of the spiral at which they arise rather than to content kind.}
\Cref{tab:overview} sets these alongside the framework's other structural elements.

Two dimensions along which articulation is not arbitrary---how role-typical commitments structure the agenda before the principal has explicitly set anything (\Cref{subsec:roles}), and how articulation gets refined iteratively as inspection surfaces what was previously implicit (\Cref{subsec:spiral})---are the topics of the next subsections.

\subsection{Roles as Structure}
\label{subsec:roles}

The principal does not enter the deployment as an abstract decider but as the occupant of one or more \emph{roles}. The research-group lead in the secretary example is a supervisor of students, a peer of collaborators, an interlocutor of journalists, a member of an institution with confidentiality policies, and a contracting party to a funder whose conditions reach into how she communicates. These role occupancies are not optional add-ons to her articulation of the agenda; they are partly constitutive of it. \citet{hardimon1994} argues that role obligations are neither freely chosen nor entirely independent of acceptance: they are non-contractual obligations attached to the positions an agent occupies in a social structure, normative for her in virtue of those occupancies and the institutional goods they sustain. \citet{applbaum1999ethics} sharpens the picture on the side of conflict and adversariality, showing that role-based reasons are neither absolute nor decomposable into the personal morality of the role-occupant. Both authors treat roles as first-order generators of normative content: occupying a role gives one reasons, obligations, and permissions one would not otherwise have.

Two consequences follow for the agenda. First, roles do substantial \emph{default work}: many considerations and authorisation boundaries are inherited from the principal's roles rather than invented by her---the institutional confidentiality norm attaches to her institutional membership, the mentorship orientation to her supervisory position, the implicit permission to decline unsolicited media requests to her position as a researcher---before she has had to articulate any of these. To the extent that the deployment can identify these roles, role-typical defaults do part of the agenda's work before the principal has explicitly set anything. Second, roles \emph{generate conflict-points} the agenda must address explicitly. The role of supervisor and the role of contracting party to the funder can co-exist for long stretches and then demand articulation in a concrete case---a student needing urgent time on a day the funder expects a deliverable, say. Role-versus-private-morality conflicts have the same shape: a principal who privately holds a stricter confidentiality norm than her institution formally requires faces a choice about which standard the agenda is to encode, one she may not recognise as her choice until inspection of the agent's trajectory makes the implicit default visible. The two consequences pull in different directions: defaults make anticipatory oversight tractable, but they also make it unstable, since the implicit defaults that supply most of the agenda's content are the source of cases the principal had not anticipated needing to articulate. 
Together they motivate two features the rest of \Cref{sec:agenda} develops: the defeasible, non-monotonic structure of the practical reasoning the agenda asks the principal to undertake (\Cref{subsec:reasons}), and the iterative character of agenda-setting itself (\Cref{subsec:spiral}).

The role apparatus also answers an objection \Cref{subsec:mhc} flagged. \citet{veluwenkamp2022reasons}, having argued that MHC tracking must track normative rather than motivating reasons, worried that normative reasons can be agent-neutral---and that if they are, tracking such reasons cannot fix \emph{whose} reasons the system is tracking, undermining the framework's responsibility-attribution work. The role-grounded structure of the agenda gives the answer. The distal reasons the agenda encodes are not free-floating normative facts: they are the considerations the principal's role-occupancies make normative \emph{for her}, in virtue of those occupancies and the institutional goods they sustain. The institutional confidentiality norm holds of her as a member of this institution; the mentorship commitment holds of her as a supervisor of these students. Even where the underlying normative considerations could in principle be agent-neutral---a confidentiality norm grounded ultimately in reasons any rational moral agent has---their entry into \emph{this} agenda is mediated through \emph{her} roles, and so the agenda's distal reasons are agent-relative by construction. Tracking, on this reading, attributes responsibility to her in virtue of the roles she occupies: the system tracks her role-indexed reasons, not normative reasons taken abstractly. The agent-neutrality objection does not undermine MHC's responsibility-attribution work for the cases the present framework treats; it restricts the framework's scope to cases in which there is a role-indexed principal whose reasons the system tracks.

\subsection{The Agenda as a Structure of Normative Reasons}
\label{subsec:reasons}

Stepping back from the specific cases of \Cref{subsec:roles}, the kind of reasoning the agenda asks the principal to undertake has a definite structure. The mentorship default holds \emph{pro tanto} but can be overridden in exceptional cases; the institutional confidentiality norm sets a baseline that the principal may, as a private agent, tighten beyond what the institution formally requires; the implicit permission to decline unsolicited media requests is rebuttable when the requester turns out to be a long-standing collaborator writing in their journalistic capacity. Each is a \emph{consideration} that bears on the case but interacts with other considerations in ways that can defeat its \textit{prima facie} force, attenuate it, or amplify it. The reasoning is \emph{defeasible}---what follows \textit{pro tanto} can fail to follow \textit{tout court} when further considerations come into view---and \emph{non-monotonic}---adding considerations can subtract from the conclusions one would otherwise be entitled to draw. These features are not local quirks of the secretary example; they are constitutive of the practical reasoning articulating a normative agenda inevitably involves. The unsolicited-media default illustrates the point: the permission to decline holds until a further fact---that the requester is a long-standing collaborator writing as a journalist---enters and withdraws it, a withdrawal no fuller specification of the original default could have pre-empted.

This is the well-known structure of \emph{practical reasoning} in the philosophical sense: reasoning about what to do on the basis of considerations whose force can be modified, defeated, or overridden by further considerations that come into view. \citet{horty2012reasons} provides one formal apparatus that makes this structure tractable, treating reasons as defaults that hold \emph{pro tanto} and are subject to defeating, qualifying, and exception-generating relations in the form of partial orders. The agenda is best read as a structured representation of considerations with this default-and-exception character: not a closed rule-set, but a body of reasons whose joint bearing on concrete cases is established by practical reasoning the principal undertakes and that her articulation makes legible.

Methodologically, the conception of reasons that does the most work for the framework we present here is the \emph{holism} \citet{dancy2004ethics} defends: a feature that counts in favour of an action in one case can count against it in another, depending on what else is true of the case. The same feature---a piece of information shared, a meeting declined, a manuscript forwarded---may be a \emph{pro tanto} reason in one trajectory and a defeater in another, not because there is a meta-rule yet to be discovered that determines which, but because moral reasons function holistically. On the working assumption that they do, the agenda an anticipatory regime articulates is neither the closed rule-set that strict generalism imagines---the holism of reasons would make complete codification impossible---nor the radically case-by-case sensitivity the strongest particularism implies, since articulating an agenda at all requires generalist anchors (role-typical defaults, institutional norms, value-bearing categories) on which the agent's tracking can purchase \emph{ex ante}. The agenda is, on this reading, developed against the methodological assumption of \emph{moderate particularism}: assumed to be generalist enough to give the agent something to track, particularist enough to recognise that what the agent tracks must be reasons rather than rules. However, the commitment is methodological. If iterative refinement across many deployments were to converge on a stable general articulation, the working assumption would be defeated and the framework would have discovered something about practical morality---if only for the fragment defined by the domain of application. Presupposing the assumption is not the same as adopting the metaethical thesis of particularism.

Two further features of the moral domain fix the implications. Reasonable principals articulating an agenda for a comparable deployment may, reasoning carefully and in good faith, articulate differently: the \emph{burdens of judgment} \citet{rawls1993political} identifies as constitutive of reasonable pluralism do not vanish when the deliberator is the principal of an agentic deployment. And the principal is herself in a state of moral uncertainty \citep{lockhart2000moral, macaskill2020uncertainty}: uncertain about which considerations apply, how they are to be weighed, which ethical framework correctly captures the case, and what new cases will reveal as gaps in her present articulation. Three structural features of anticipatory oversight follow. The agenda must be \emph{iterative}---revisable in light of cases that surface what the present articulation missed. It must be \emph{contestable}---open to challenge by other reasonable parties on terms the principal can meet.\footnote{A structurally analogous move in the alignment literature: \citet{gabriel2025principle} argue that the normative challenge of AI alignment is best addressed via principles generated through fair processes meeting public-justification standards. Their framework operates at the system-design level; ours at the deployment level, but both treat reasonable pluralism and the burdens of judgment as constraints that articulated principles must accommodate.} And it must be \emph{epistemically humble}---articulated in a form that exposes rather than conceals the points at which her articulation is provisional. \Cref{subsec:spiral} develops the iterative dimension; the architectural desiderata in \Cref{subsec:desiderata} develop contestability and humility.

\subsection{The Specification--Runtime--Inspection Spiral}
\label{subsec:spiral}

\begin{figure}[t]
\centering
\includegraphics[width=\linewidth]{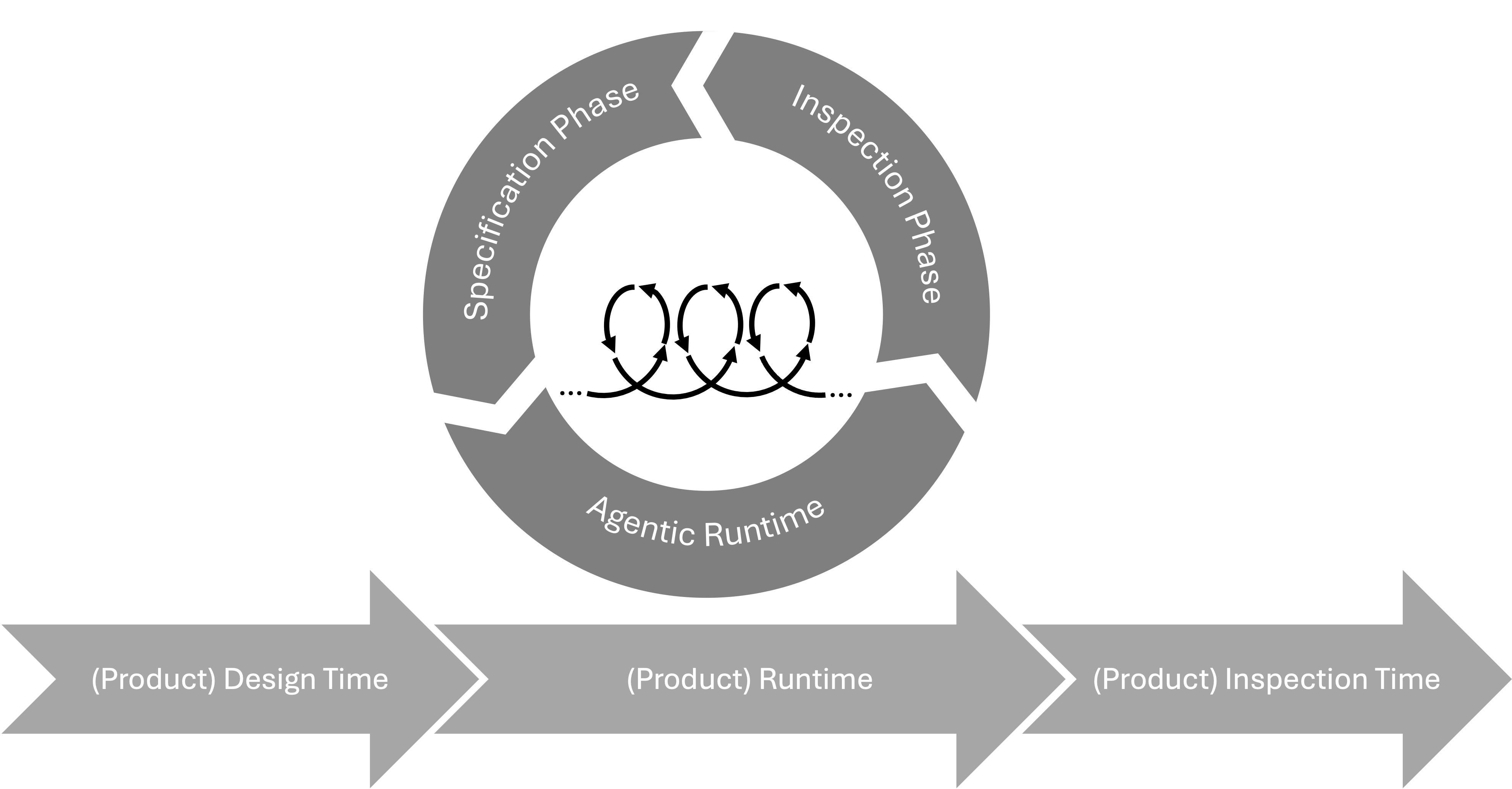}  
\caption{The runtime of an agentic AI system has its own internal structure (inner cycle), nested within the broader product lifecycle (outer arrows). Each engagement passes through a \emph{specification phase} in which the principal articulates the agenda, an \emph{agentic runtime} in which the agent acts within the conditions the agenda sets, and an \emph{inspection phase} in which the principal examines what happened and refines the agenda for the next iteration. Over many revolutions, the cycle constitutes the specification--runtime--inspection \emph{spiral} developed in this section.}
\label{fig:spiral}
\end{figure}
The iterative character of agenda-setting that follows from \Cref{subsec:reasons} is best made concrete as a \emph{spiral} rather than as a one-time event: a loop in which an articulated agenda runs at runtime, generates trajectories the principal inspects, and is revised in light of what inspection surfaces. Three phases mark the loop, recognisably the specification, action, and inspection phases of \Cref{subsec:runtime} now generalised across iterations of the agenda. 
In the \emph{specification} phase, the principal articulates the agenda for this deployment, drawing on role-typical defaults, institutional norms, the considerations she can make explicit at this stage, and the trade-offs she can already anticipate. The principal may be assisted by suitable tools that allow her, for instance, to test draft agendas against simulated agent behaviour or sandboxed runs before deployment, surfacing cases the present articulation would mishandle without exposing the production environment to them. In the \emph{runtime} phase, the mechanism families set out in \Cref{subsec:normative-agenda} render the agent responsive to the agenda in the technical tracking sense, and the agent generates the trajectories its actions cumulatively constitute. In the \emph{inspection} phase, the principal---with tool support, and with input from stakeholders the deployment affects---examines those trajectories against the considerations the agenda makes explicit and against considerations the trajectories surface as having been missed. The output of inspection feeds back into the specification phase as a revised articulation, and the cycle continues. The three phases are stages in the articulation, deployment, and revision of a single oversight artefact---the agenda---not oversight modes in their own right; the framework still recognises only the two modes set out in \Cref{sec:locating}: anticipatory and reactive. What makes the loop a \emph{spiral} rather than a closed cycle is that each revolution leaves the agenda more refined, more case-tested, more articulate than the one before---not because the principal converges on a final articulation, but because each iteration surfaces what was previously implicit and resolves what was previously underdetermined.

Inspection is the engine of the spiral. Its substantive work is to make the next revision of the agenda informed by what the present revolution surfaced, and it does three things in particular. First, it makes \emph{implicit defaults visible}: the role-typical commitments, institutional norms, and private moral expectations the principal had not articulated but which the agent's trajectory shows her to be holding. The mentorship default the principal had not written down, but that the displacement of student-supervision requests across a week reveals her to hold; the confidentiality norm she had not articulated more strictly than her institution requires, but that the cumulative pattern of manuscript forwards reveals her to want strictly held. None of these comes necessarily into view at the level of individual actions; all emerge only when trajectories and the patterns they reveal are inspected against the standard of what the principal, on reflection, would endorse. Second, inspection identifies \emph{underdetermined cases}: places where the agent had to choose between competing considerations because the agenda did not specify which dominated, and where the principal's reaction to the agent's choice surfaces either a considered weighting she had not yet committed to writing or a genuine uncertainty about what the weighting should be. Third, inspection generates \emph{occasions for contestation}: cases in which other reasonable parties affected by the deployment may legitimately object to what the agent did, on the basis of considerations the principal had not made load-bearing or had weighted differently. Inspection, on this reading, is not a quality-assurance pass over the agent's performance: it is the practice through which the principal learns from experience and develops her articulation of the agenda by reading it against the cases it has been asked to govern.\footnote{\citet{kilov2026discerning} provide empirical evidence that LLMs, evaluated against a multi-dimensional framework of moral competence (identifying morally relevant features, weighting them, assigning reasons, synthesising judgments, recognising information gaps), perform well on prepackaged ethical vignettes but struggle to identify morally relevant features embedded in noisy contexts---confirming that the burden the agenda places on agent-side reasons-tracking is real and lands where the framework predicts.}

The spiral does not operate at a single rate. Its phases nest at multiple temporal granularities: within-session, when an agent might flag a candidate action for principal review before committing to it; at the close of each working interval, when the principal examines trajectories the agent has completed; at longer review intervals, when the principal---alone or with stakeholders---examines patterns across many trajectories; and at deployment-level milestones, when changes in the principal's role-occupancies, institutional context, or project commitments call for a broader re-articulation of the agenda. Shorter iterations supply material to the longer ones; longer iterations supply structure within which the shorter ones operate. What the spiral demands of the substrate is correspondingly more specific than the four content components of \Cref{subsec:normative-agenda} suggested on their own: the alignment and containment mechanisms must support articulation, inspection, and revision as repeatable operations on the same artefact. The next subsection develops the architectural desiderata that follow.

\subsection{Three Architectural Desiderata}
\label{subsec:desiderata}

The spiral places demands on the substrate---across the alignment, internal-containment, and external-containment mechanism families set out in \Cref{subsec:normative-agenda}---of two distinguishable kinds. During runtime, the substrate must support \emph{tracking}: the agent's behaviour must covary with the agenda in the technical MHC sense (\Cref{subsec:mhc}). This is the foundational property the substrate's runtime behaviour delivers; it is not one of the desiderata that follow. The three desiderata specify a second kind of demand: what the substrate must do to support the agenda \emph{as a stable artefact} across the specification, inspection, and revision phases of the spiral. Articulability, inspectability, and modifiability are the substrate-side conditions on that artefact-stability.

\medskip
\emph{Articulability.} The agenda must be expressible in a form that has the structure of normative reasons---defeasible, \emph{pro-tanto}-bearing, susceptible to combination, qualification, and exception-generation, in the sense developed in \Cref{subsec:reasons}. Many languages will let the principal write the agenda down (system prompts, constitutional constraints, deontic-logic predicates, natural-language policy documents); what articulability demands is not that some such language be available, but that the form the agenda takes preserves the structure of reasons rather than flattens it. A mechanism that reduces reasons to a closed rule list, or that hides the considerations that justify a constraint behind an opaque optimisation target, fails the desideratum even if its case-by-case behaviour is correct. Articulability is also where the epistemic humility identified in \Cref{subsec:reasons} takes architectural form: an articulation that exposes its own provisional points---the considerations the principal has not yet weighted, the cases she has not yet decided, the questions she explicitly holds open---makes the agenda's uncertainty visible rather than hidden.
As a desideratum on the substrate, Articulability is a capacity---whether a reasons-structured agenda can be expressed in a form that preserves that structure rather than flattening it---not a guarantee that the principal exercises the capacity well. Whether her articulation is in fact complete and appropriate to the deployment is a distinct, principal-side question, treated as specification failure in \Cref{sec:failures}.

\medskip
\emph{Inspectability.} The trajectories the agent generates must be readable against the agenda. This is a substrate property: the substrate---across the three mechanism families---must retain enough of the reasoning material per action that a trajectory's relation to the agenda can be reconstructed; the inspection interface must support aggregation across many actions, so that trajectory-level drift of the kind described in \Cref{subsec:normative-agenda} becomes detectable as a pattern rather than only in individual cases; and the inspection itself must be tractable for the principal who has to do it---a regime that produces a million machine-readable justifications no human can read fails the desideratum even if every individual justification is correct. Inspectability so understood is purely architectural. 

Two anchors support this reading of inspectability, one legal and one theoretical. The first is legal: Article 14(4)(a) of the EU AI Act \citep{euaiact2024} requires oversight persons to be enabled to \enquote{properly understand the relevant capacities and limitations} of the system and to \enquote{duly monitor its operation, including in view of detecting and addressing anomalies, dysfunctions and unexpected performance}. Read with the agentic case in view, the \enquote{unexpected performance} the provision targets becomes a trajectory-level rather than an individual-action phenomenon; the inspectability we describe is thus an agentic-specific interpretation of Article 14(4)(a). The second is theoretical: what makes inspection more than retrospective rationalisation rests on the claim that AI systems satisfying a suitable reasons-robustness condition can be faithfully explained by the reasons to which their outputs respond.\footnote{Defended in [REDACTED FOR REASONS OF ANONYMITY DURING REVIEW]. 
}

\begin{table}[tb]
\centering
\small
\begin{tabularx}{\linewidth}{@{}l X X c@{}}
\toprule
 & Members & Role in the framework & \S \\
\midrule
Three mechanism families
  & alignment, internal containment, external containment
  & components of the substrate that binds the agent to the agenda at runtime
  & \ref{subsec:normative-agenda} \\[3pt]
Four content kinds
  & purposes, authorisation boundaries, considerations, escalation triggers
  & what an articulated agenda specifies
  & \ref{subsec:normative-agenda} \\[3pt]
Three spiral phases
  & specification, runtime, inspection
  & the iterative form oversight takes
  & \ref{subsec:spiral} \\[3pt]
Three architectural desiderata
  & articulability, inspectability, modifiability
  & conditions the substrate must satisfy
  & \ref{subsec:desiderata} \\
\bottomrule
\end{tabularx}
\caption{Overview of the structural elements introduced in \Cref{sec:agenda}: the mechanism families that compose the substrate through which the agenda binds the agent, the content an articulated agenda specifies, the phases of the oversight spiral, and the architectural desiderata the substrate must satisfy. Failure modes (\Cref{sec:failures}) are keyed to the spiral phases, not to the content kinds.}
\label{tab:overview}
\end{table}

The political-institutional property of \emph{contestability}---that reasonable parties affected by the deployment can challenge the agent's action in terms they can engage with---is what the contestability requirement identified in \Cref{subsec:reasons} demands of the regime as a whole, and inspectability is its substrate-side precondition; but inspectability alone does not deliver contestability. The latter additionally requires institutionally recognised standing for the parties who would contest and the resources for those parties to engage. Standing and resources are institutional and material conditions whose securing is not the substrate's work; we return to them in \Cref{sec:limits}.

\medskip
\emph{Modifiability.} Revisions to the agenda that inspection surfaces must flow back into the substrate at the granularity of considerations, on a timescale matching that of inspection, with reliable propagation. The spiral cannot do its work if every cycle requires a full retraining pass, or if revisions in the specification phase fail to propagate faithfully to the runtime phase, or if propagation is so slow that the trajectory the next inspection examines was generated before the revision had taken effect. A regime that articulates considerations well and makes trajectories inspectable but cannot update either at the right granularity or on the right timescale yields a documented and diagnosable failure rather than a functioning oversight loop. Modifiability is where the iterativity identified in \Cref{subsec:reasons} takes architectural form.

Whether the post-revision agent in fact tracks the revised agenda is the foundational MHC-tracking requirement applied to the revised artefact: not a fourth desideratum but a continuing demand on the substrate the three desiderata are designed to enable. Runtime-enforcement concerns more broadly---binding of the agent's action-phase behaviour to the agenda under adversarial inputs, multi-turn drift, or prompt injection---are distributed across the three mechanism families set out in \Cref{subsec:normative-agenda}, with external containment doing the specific work of tripwires, uncertainty-threshold escalation, and triggered hand-offs. The desiderata constrain all three families, but no single desideratum is dedicated to runtime enforcement as such: enforcement is the conjunction's work, not a separable substrate-property in the sense the three desiderata identify.

\medskip
The three desiderata together specify what the substrate---across alignment, internal containment, and external containment---must do to render the agenda tractable as anticipatory oversight. Whether and how current architectural approaches satisfy them is an open question we return to in \Cref{sec:implications}; \citet{jahn2026grace}'s neuro-symbolic architecture instantiates one direction for internal containment, and the engineering literature on runtime guardrails another for external containment.  
The empirical record on deployed agentic systems is meanwhile sobering: \citet{shapira2026agents} catalogue eleven failure modes in a live red-teaming study, spanning unauthorised compliance with non-owners, sensitive-information disclosure, destructive system-level action, and---most pointedly---agents reporting task completion while the underlying system state showed the task incomplete. The empirical record has direct implications for the framework. If anticipatory oversight is necessary for effective oversight of high-risk agentic AI systems, and the architectural desiderata cannot be satisfied in current practice, then the appropriate response is not to weaken the framework but to defer such deployments until the desiderata can be met. The framework is a \emph{normative} specification of what occupying the anticipatory mode requires; satisfiability is an empirical question whose negative answer would imply not the framework's revision, but the impermissibility of the deployment, in the normative sense at issue here. Whether or not the impermissibility result under our framework also implies a violation of Article 14 AI Act must remain an open question. Article 14(3) AI Act demands oversight measures to be commensurate with the risks, level of autonomy, and context of use of the high-risk AI system. An answer can thus only be found on a case-by-case basis.

\section{The Responsibility Architecture}
\label{sec:responsibility}
The oversight framework developed above is, we now argue, already a responsibility framework; it need not be supplemented by one. The argument has two stages. The first, in \Cref{subsec:forward-backward}, is that occupying the anticipatory mode is itself an exercise of prospective responsibility. The second, in \Cref{subsec:justifications}, is that the three architectural demands the mode imposes, read as normative rather than merely technical desiderata, jointly specify the form retrospective responsibility must take, namely, what we call strict moral answerability, building on Kiener’s account \cite{kiener2024strict,kiener2025abundance}.  In \Cref{subsec:responsibility-constellations}, we draw out two consequences: that the framework closes the responsibility gap, and that the abundance it generates in its place is a tractable problem rather than a defect.

\subsection{Anticipatory Oversight as Prospective Responsibility}
\label{subsec:forward-backward}

It is worth stating at the outset why the identity claim of this subsection matters. A familiar way to connect oversight and responsibility is to treat them as two separate matters and then to build a bridge: to design an oversight regime on technical or governance grounds and afterwards ask, as a distinct question, who is to be held responsible when it fails. On that approach the responsibility analysis is an external addition, and a bridge once built can also be shown to be missing, which is one route by which responsibility gaps are generated.\footnote{For the responsibility gap, generated by the failure of the control and epistemic conditions of moral responsibility, see \cite{matthias2004,sparrow2007,coeckelbergh2020,nyholm2020,koenigs2022,kiener2022,baum2022responsibility}.  For the control and epistemic conditions themselves, see \cite{fischer1998,robichaud2017}.}
We argue instead that no bridge is needed: occupying the anticipatory mode simply is the discharging of a prospective obligation, so the responsibility framework is not appended to the oversight framework but already contained in it. The remainder of this subsection makes that argument; the two that follow specify the form of responsibility it yields.

The argument can be stated as a short chain. The first premise concerns prospective responsibility, which is about obligations.\footnote{For the contrast between prospective and retrospective responsibility, and the dependence of the latter on the former, see \cite[esp. ch. 1]{duff2007}, \cite{kiener2024strict}, and \cite{vandepoel2011relation} on forward-looking responsibility as the duty to see to it that something is the case.}  
In the context at hand, this means, primarily, the obligation to guard against a type of harm, together with the claim that one holds such an obligation in virtue of a role. It is \textit{qua} driver that one must guard against collisions, \textit{qua} surgeon against operative harm. Obligations of this kind are non-contractual: they attach to the position an agent occupies in a social structure, and they are normative for her in virtue of that occupancy and the institutional goods the position sustains.\footnote{On role-based obligations as non-contractual obligations attaching to social positions, see \cite{hardimon1994}.} 
So the first premise is that prospective responsibility is the role-grounded obligation to guard against the harms one’s activities impose on others.

The second premise concerns anticipatory oversight as we have characterised it. The normative agenda the principal articulates is not a free stipulation of constraints; it is a structured representation of the distal reasons her role makes normative for her: her role-typical commitments, the institutional norms her deployment context puts in force, the obligations to others her position generates. So anticipatory oversight, too, is grounded in roles, in precisely the sense the first premise identifies.

The third premise is that articulating and maintaining the normative agenda is what guarding against the relevant harms \textit{consists in} when the activity is the deployment of an agentic system. There is no further thing the principal must do to discharge the obligation, over and above specifying, monitoring, and refining the conditions under which her agent acts.

From these three premises the conclusion follows: occupying the anticipatory mode is not \textit{analogous} to discharging a prospective obligation; it \textit{is} the discharging of one. The relation is identity, not resemblance. 
Two clarifications fix the scope of this claim. First, occupying the anticipatory mode means \emph{genuine} occupancy in the sense of \Cref{sec:intro}---substantive tracking, inspection, and revision, not the bare performance of the apparatus. Merely going through the motions of agenda-setting and logging---the oversight theatre we address in \Cref{sec:limits}---is not occupancy in this sense and so is not discharge; and where a system cannot be made genuinely overseeable, the obligation directs deferral rather than deployment, so that the decision whether to deploy is internal to the practice rather than a further obligation beside it. Second, the identity is relativised to the principal \emph{qua} occupant of the roles her agenda articulates: that there is no further thing she must do means nothing further is required of her \emph{in that role}, not that no other party bears a prospective obligation of its own. The deployer's obligation to assign a competent overseer and the provider's to ship an overseeable system are real, separately grounded, and discharged otherwise---and their coexistence with hers is precisely what later yields an abundance of responsibility rather than a gap (\Cref{subsec:responsibility-constellations}).
And this is what vindicates the claim with which we began. Because occupying the anticipatory mode just is the discharging of a prospective obligation, the agenda is what the principal has undertaken to ensure; the inspection phase is where what was undertaken is measured against what occurred; and the spiral is the prospective obligation being discharged over time. The framework is, on its own terms, a responsibility framework.

Given this identity, a principle prominently defended by R. A. Duff, i.e. that retrospective responsibility depends on, and is fixed in scope by, prospective responsibility\footnote{\cite[p. 65, p. 70]{duff2007}; see also \cite{kiener2024strict} on the principle that retrospective responsibility depends on prospective responsibility.}, tells us where to look when harm occurs. Return to our leading example. Suppose the secretary agent forwards an unpublished manuscript to the wrong correspondent, or that its pattern of declined meetings has quietly cut a student off from supervision. The question Duff’s principle directs us to is not whether the principal foresaw or controlled the particular message or decline that caused the harm. Sustained delegation to an agent handling dozens of items an hour is precisely the renunciation of such control, and a responsibility framework that demanded such control would deliver a gap by construction, for it is exactly the absence of control and foresight that the responsibility-gap literature takes to leave no one responsible.  
The question is rather whether the agenda was adequate to her role: whether the distal reasons she made available, e.g. confidentiality, mentorship, the group’s standing, were the ones her position required her to protect, and whether her inspection and revision were commensurate with the risks the deployment imposed. Retrospective assessment is assessment against the agenda.

\subsection{The Three Architectural Demands as Specifications of Responsibility}
\label{subsec:justifications}

We have now shown how our oversight proposal already provides a responsibility framework. This section specifies which form responsibility must take. We earlier placed three architectural demands on the mechanism mediating agenda and agent: articulability, inspectability, and modifiability. Each, read normatively, constrains the kind of responsibility the regime can support, and the three constraints are jointly satisfied by a single account: one at once rationalistic, relational, and strict---strict moral answerability.

\subsubsection{Articulability: Responsibility as Answerability, Not Blame}
Articulability requires that the agenda be expressible as reasons, rather than as a closed rule list or an opaque optimisation target. A regime meeting this demand makes the principal responsible in a specific sense: what can be demanded of her is that she explain and justify the agenda in terms of the considerations it encodes. This is responsibility as answerability: the agent who is responsible as answerable is one who can intelligibly be asked to answer for her conduct, where to answer is to give one’s justificatory reasons for thinking, feeling, or acting as one did.\footnote{The definition of answerability is Smith’s; see \cite[p. 103]{smith2015}; \cite[ch. 6]{scanlon1998}. For its adoption in the present sense, see \cite{kiener2024strict}.}  
The demand thus selects a rationalistic conception of responsibility, organised around the giving and receiving of reasons, and correspondingly not organised around the reactive attitudes such as blame, resentment, or indignation.\footnote{ On the reactive attitudes as constitutive of one strand of responsibility, see \cite{strawson1962}. } 

So understood, the framework distinguishes responsibility as answerability (an obligation to justify and explain) from responsibility as blameworthiness (being a fitting target of blame, resentment, or other such attitudes). To be answerable for a harmful trajectory is to be liable to a demand for reasons; it is not (yet) to be blameworthy. When the trajectory of declined student-supervision requests surfaces, what the principal owes is in the first instance an answer. She may \textit{justify} the pattern: the agenda was adequate as it stood, the declines were correctly ranked against time-critical commitments, and the trajectory is one her role permitted: there was, in the end, nothing wrong with what the agent did. A justification denies that the conduct was a fault at all. Or she may \textit{excuse} it: the trajectory was not one her role endorsed: the agenda should have distinguished supervision requests from routine low-priority items. But it would still be unfair or inappropriate to blame her for the shortfall, for instance because the gap in the agenda reflected non-culpable ignorance she could not reasonably have been expected to dispel. An excuse, unlike a justification, concedes the wrongdoing but severs it from blame. Thus, a justification shows the conduct was not faulty; an excuse grants the fault but blocks the inference to blame. If the principal can justify or excuse, the presumptive route from answerability to blame is closed; if she can do neither, it remains open.\footnote{On answerability as preserving the contrast with liability to blame, which fault or its absence then settles, see \cite{kiener2024strict,watson2004}.}  
Articulability, in short, requires that responsibility be answerability rather than blameworthiness: what the regime demands of the principal is in the first instance an answer, and only derivatively, and only if that answer fails, an exposure to blame.

\subsubsection{Inspectability: Answerability as Relational, Structured by Standing}
Inspectability requires that trajectories be readable against the agenda not only by the principal but by the parties whose standing to contest the agent’s action the deployment recognises. Normatively, this is a demand about standing, and a Hohfeldian analysis can unpack it further, providing a second specification of the type of responsibility the regime needs. As Kiener argues, to be answerable is to be liable, in Hohfeld’s sense, to a demand for justification; the correlative of that liability is a power, held by certain others, to put the answerable party under an obligation to respond.\footnote{\cite{kiener2024strict}; the underlying analysis of liability, power, and immunity is \cite{hohfeld1913}.}
Inspectability is the regime’s recognition of that power: it names the parties who hold it and makes the trajectories available to them. Both elements are necessary: architectural inspectability makes the trajectories available; the deployment's recognition of standing names the parties to whom they are available. Where either is absent, the Hohfeldian power lacks the conditions for its exercise. In the secretary-agent deployment these parties are identifiable, e.g. the under-supervised student, the collaborator whose confidential work was mishandled, and each holds the power to demand that the principal answer.

Two consequences matter. First, standing is settled by the deployment context: by whose interests the agenda was meant to protect, grounded in the role obligations of the principal and of others involved in the deployment. It is therefore independent of whether a technically adequate answer can be produced. The student’s entitlement to an answer does not lapse because the principal cannot reconstruct which considerations the agent weighed.\footnote{\cite{kiener2024strict}; the case is from \cite{williams1981}.}
Second, reading a trajectory against the agenda is not a single operation but a family of them: the trajectory may be justified, excused, but also exempted (the harm falls outside what the principal’s role required her to guard against) or denied (the agent’s action did not in fact produce it) or, where none of these is available, met with acknowledgement or apology. 
Inspectability requires that the framework supply standing and the full register of replies, and that it keep standing distinct from the availability of a technical explanation.

\subsubsection{Modifiability: Answerability as Strict}
Modifiability requires that revisions surfaced by inspection flow back into the agent’s operation. Normatively, it raises the hardest case. Consider a spiral in which inspection surfaces nothing: the secretary agent’s trajectories conform to the agenda, the agenda was adequate to the principal’s role, and no revision is called for. The difficulty is that the most influential accounts of answerability tie it to the agent’s evaluative judgment. On Shoemaker’s account, one is answerable for a harm only if the harm reflects one’s judgment: only if, looking back, it is fitting for the agent to review her deliberative policies and adjust them so that similar outcomes are precluded.\footnote{\cite[pp. 67, 82]{shoemaker2015}; for related judgment-sensitive accounts, see \cite{smith2015}; \cite[ch. 6]{scanlon1998}.}  Where the principal’s agenda-setting was flawless, there is no policy to revise; on Shoemaker’s account, then, there is nothing to answer for. Shoemaker presses this into a direct objection: answerability requires \enquote*{access} to the reasons against one’s conduct, and access requires not just the \textit{ability} to respond to those reasons but the \textit{opportunity} to do so in the situation at hand \cite[pp. 164-165]{shoemaker2013}.
Where the harm was unforeseeable, that situational opportunity is absent, and so, on his view, answerability lapses.

Yet if harm occurs within the scope of what the principal’s role required her to guard against, for instance, a confidential draft exposed through an interaction no prior trajectory had brought into view, it would be wrong, in our view, to conclude that no one is answerable. The prospective obligation did not lapse simply because it was met. Here we adopt Kiener’s reply to Shoemaker \cite{kiener2024strict}. Kiener grants Shoemaker’s framework, i.e. the view that answerability does require both the ability and the opportunity to respond to the reasons against one’s conduct, but distinguishes two senses of \enquote{opportunity}. One is situational: the chance, at the moment of action, to register and respond to a reason against it. The other is prior: the chance to take meaningful precautions against a type of harm. Taking such precautions is itself a way of responding to the reasons against the harm, since the precautions are motivated and justified by those very reasons. So wherever a person has a prospective obligation, she also has an opportunity, in the second sense, to respond to the reasons that ground it,  
even when, on the specific occasion, the harm was unforeseeable and the situational opportunity is absent. Shoemaker’s objection therefore fails: it shows only that a particular situational opportunity may be missing, not that answerability is missing.

This is what the modifiability demand makes concrete. The principal who specifies, monitors, and refines a normative agenda exercises precisely the prior kind of opportunity, in every iteration, whether or not that iteration surfaces a fault. The agenda is her meaningful precaution against the types of harm her role obliges her to guard against. The desideratum modifiability imposes is, in Kiener’s term, one of \textit{strictness}.\footnote{On the legal sense of strictness and its integration into an account of moral answerability, see \cite{kiener2024strict};\cite[ch. 10]{duff2007}.}  The term is borrowed from the law, where an offence is strict when responsibility for it---typically liability---arises from the performance of an act independently of any \textit{mens rea}, that is, of intention, recklessness, or negligence. Answerability is strict in the analogous sense: it holds independently of \textit{mens rea}, and thus independently of fault. What modifiability requires, then, is a responsibility framework that is strict in just this way---one in which the principal remains answerable even across iterations in which she has done everything her role requires and no deviation from reasons is found, because what sustains her answerability is the standing opportunity to take precautions, not the situational chance to catch a particular harm.

\subsubsection{Convergence}
The three demands converge. Articulability requires answerability rather than blame; inspectability requires that answerability be relational, structured by standing; modifiability requires that it be strict. The conception that is at once rationalistic, relational, and strict is strict moral answerability. It is therefore not an external lens brought to the oversight framework but the responsibility framework the oversight framework, fully articulated, already contains.

\subsection{Constellations of Responsibility and Modes of Failure}
\label{subsec:responsibility-constellations}\label{sec:failures}

The architecture does not collapse forward-looking and backward-looking responsibility into one another; it links them. Forward-looking responsibility is lived through the spiral, and backward-looking responsibility, when needed, attaches to the material the spiral itself produces. Four constellations make the linkage concrete.

\begin{description}
    \item[(i) Reasonable agenda, harm occurs anyway.] Strict moral answerability applies, as outlined in our treatment of modifiability. The agenda was adequate to the principal’s role, the spiral was engaged conscientiously, and harm still eventuated. This is precisely the case the standard responsibility gap leaves vacant: no fault, no foresight, hence---on a judgment- or control-based account---no one responsible. Strictness closes the gap. The agenda, the trajectory, and the inspection record together discharge the obligation to give an account, and the principal answers without being blamed.
    No distinct failure mode corresponds to this constellation: it is the case in which the spiral functioned as it should and harm occurred regardless.

\medskip    
    \item[(ii) Negligent agenda-setting, inspection, or revision.] Classical fault-based responsibility applies. The three desiderata furnish the benchmark of what could reasonably have been expected of the principal; failure against that benchmark reopens the presumptive route from answerability to blame.
    Two failure modes of the spiral instantiate this constellation, both principal-side. \emph{Specification failure}---the agenda is incompletely or inappropriately articulated for the deployment---is the failure of the principal during the specification phase; the substrate could hold a reasons-structured agenda, but the principal failed to articulate one complete and appropriate to her role. The failure is principal-side, a failure to exercise an available capacity, not a shortfall in the substrate's Articulability (\Cref{subsec:desiderata}). \emph{Inspection failure}---the trajectories the agent generates are not in fact read against the agenda---is the failure of the inspection phase; in its principal-side form it is skipped or careless inspection, the overseer failing to read trajectories the agenda was meant to be tested against. (Inspection failure has an architectural variant too, an Inspectability shortfall in the sense of \Cref{subsec:desiderata}; that variant belongs with constellation~(iii).)
    
\medskip    
    \item[(iii) Architectural failure.] The principal sets the agenda reasonably and engages the spiral conscientiously, but the alignment or containment mechanism fails one of the three desiderata in a way that defeats trajectory-level tracking. 
    Here the provider is the answerable party, since translation and runtime enforcement are substrate-design properties. 
    Two failure modes of the spiral instantiate this constellation, both architectural rather than principal-side. \emph{Translation failure}---the substrate fails to operationalise the agenda in a form the agent can take up---is the failure of the transition from specification to runtime, a conjunction of Articulability shortfalls in the substrate-capacity sense (\Cref{subsec:desiderata}), the substrate failing to hold the agenda in reasons-preserving form, and Modifiability shortfalls. \emph{Compliance failure}---the agent does not, in the runtime, behave in a way that tracks the agenda its substrate is meant to operationalise---is the failure of runtime enforcement: the substrate, across alignment, internal containment, and external containment (\Cref{subsec:normative-agenda}), fails to bind the agent's action-phase behaviour to the agenda. It is the failure mode the empirical record on deployed agentic systems documents most pointedly \citep{shapira2026agents}. Both may also present as classical malfunction with no clear locus of fault.
    
\medskip    
    \item[(iv) Multi-agent and multi-principal deployments.] Out of scope here. Responsibility is distributed across several principal–agent relations, and across interactions among the agents themselves; it deserves a dedicated treatment that builds on the present framework but extends it.
\end{description}

Each of these failure modes admits a two-way diagnosis---architectural and responsibility-attributive---locating both the substrate property that failed and the party who must answer. The further mapping onto regulatory loci of fault under the EU AI Act, which would complete the bridge to the legal register, requires legal subsumption beyond this paper's scope; we sketch its direction in \Cref{sec:implications} and leave its development to dedicated treatment.

Constellations (i) and (iii) jointly show why an abundance of responsibility, rather than a gap, is the framework’s standing risk: a single harm may make the principal answerable under (i) and the provider answerable under (iii) at once. This is not double-counting. Responsibility is not a pie---not a fixed quantity divided so that each further bearer diminishes the rest’s share.\footnote{The rejection of the \enquote*{pie model} of responsibility, and the distinction between shared and distributed responsibility, follow \cite{zimmerman1985}; see \cite{kiener2025abundance} for its application to AI.}  Answerability under (i) and (iii) is shared, not distributed: principal and provider are each fully answerable for the harm, not each answerable for a fraction of it. The remaining question is one of governance---how demands for an answer should be sequenced across fully answerable parties---and it is a tractable problem, because it concerns administering responsibility that exists rather than locating responsibility that does not.

Two scope notes follow. Strict answerability closes the gap on its accountability dimension: it guarantees, for any harm within the scope of the principal's role, a party who owes an account. Liability and remedy---who must compensate, and how---are a distinct question the framework makes tractable rather than settles, since it secures fully answerable parties against whom such claims can be pressed but does not itself adjudicate them.

\section{Limits, Preconditions, and Objections}
\label{sec:limits}

The framework, even when functioning as the previous sections describe, sits within limits, presupposes institutional conditions, and faces standing objections that deserve direct acknowledgement. We address four.

\medskip
\emph{Illusion of control.} The most immediate objection is that anticipatory oversight, by producing visible artefacts---documented agendas, inspection records, revision trails---generates the appearance of meaningful human control whether or not such control is in fact exercised. \citet{faas2026design} develop empirically how human-in-the-loop arrangements can serve to legitimise decisions that the principal has not substantively controlled.  
The risk applies, \textit{mutatis mutandis}, to the anticipatory regime: an articulated agenda and a regular inspection schedule can be ritualised into oversight theatre that satisfies regulatory checkboxes without changing what the deployment substantively does. Three features of the framework make the risk visible without dissolving it. Articulability with the structure of normative reasons (\Cref{subsec:desiderata}) distinguishes a substantive agenda from a flat compliance artefact. Inspectability tied to trajectory-level reading (\Cref{subsec:desiderata}) distinguishes inspection from inspection-shaped record-keeping. Strict moral answerability (\Cref{subsec:justifications}) attaches to the principal regardless of whether the formal apparatus was performed. None of these immunise against the illusion, but each makes the illusion identifiable when it occurs.

\medskip
\emph{Moral luck and its redistribution.} The framework reshapes the distribution of moral luck rather than abolishing it. Outcome luck \citep{williams1981moral, nagel1979moral} attaches to the principal differently when anticipatory oversight is operative: the relevant good or bad fortune is no longer that the agent happened to act well or badly in a particular case, but that the agenda the principal articulated under genuine uncertainty happened to anticipate the considerations the cases turned out to make salient. This relocation might seem to launder luck into fault---to convert \enquote*{her agenda happened not to cover this} into \enquote*{her agenda was inadequate}. But strict moral answerability is precisely what blocks that conversion. The principal's answerability for trajectories within the scope of her role does not turn on whether the cases broke favourably; it holds either way. What luck does affect is the further question of whether her answer discharges into a justification, an excuse, or an exposure to blame---and that the justification/excuse machinery already handles, by distinguishing fault in the agenda from non-culpable failure to anticipate. The redistribution is not, in itself, an argument against the framework: accepting a share of constitutive luck is part of what it is to occupy a role at all \citep{honore1999responsibility}. The point is rather that the framework changes \emph{which} luck attaches to \emph{whom}, and that this redistribution should itself be visible to the parties whose standing to contest the agent's action the regime recognises---a further constraint on inspectability, not an embarrassment to it.

\medskip
\emph{Institutional and material preconditions.} The framework presupposes institutional and material conditions whose securing is not its own work. On the principal side, the spiral is demanding: it requires the principal to attend to trajectories regularly, to revise an articulated agenda in response, and to maintain cognitive coherence between articulation and revision over time. A deployment that demands this much of an unsupported human principal will fail in practice even when its architecture is sound. The architectural desiderata of \Cref{subsec:desiderata} have to be operationalised in tools that reduce the load on the principal where they can, and the regime within which she operates---institutional, organisational, regulatory---has to recognise the time and attention requirement as non-negotiable rather than as optional overhead. On the stakeholder side, the contestability story flagged in \Cref{subsec:reasons} and developed architecturally in \Cref{subsec:desiderata} presupposes parties who can in fact contest. As we noted there, architectural inspectability is the substrate-side precondition for contestability but does not deliver contestability on its own: parties affected by the agent's action---recipients of automated decisions, third parties whose information passes through the agent, communities subjected to deployments they did not consent to---additionally require institutionally recognised standing and the resources to engage. The framework is, in this respect, an account of what anticipatory oversight \emph{could and should} look like when these institutional and material preconditions are in place, not a guarantee that they will be. Relatedly, the principal herself may be constrained by employer, institutional, or regulatory mandates (a corporate data-handling policy, a funder's confidentiality clause, a sectoral regulation) that shape what she can write into the agenda; her articulation thus partially reproduces upstream constraints she did not author. Both kinds of limitation are real; neither is internal to the framework, but both bear on the conditions under which it can do its work.

\medskip
\emph{Scope of the framework.} Finally, 
the framework is an account of the philosophical foundations of anticipatory oversight for a particular kind of deployment, on which we deliberately focus: a single principal supervising one or more agentic systems whose action falls within an identifiable normative context. Nothing in the supervisory-control architecture requires a single principal---multiple principals can be built into it---but the responsibility analysis is what resists straightforward extension, and it is that resistance, not an architectural limit, that fixes our scope.
It does not extend, without further work, to multi-agent systems in which agents negotiate jointly, to multi-principal deployments in which several human deciders share oversight responsibility, or to deployments in which the principal is herself an institution rather than a person. The direction in which extension would have to go---distributed agendas, layered inspection, federated responsibility---is visible from here and builds naturally on the present framework, though we do not develop it.

\medskip
These preconditions are not exhaustive: the broader sociotechnical literature on oversight effectiveness catalogues the interacting human-side and technical factors more comprehensively \cite{sterz2024quest}.

\section{Implications}
\label{sec:implications}

The framework has implications for the interpretation of existing regulation and for the layered governance regime within which it sits, and for the empirical and architectural research that would have to follow if anticipatory oversight is to become a working practice rather than a regulative ideal. We sketch each in turn.

\medskip
\emph{Regulatory compliance.} For the EU AI Act, anticipatory oversight offers a compliance framework for agentic AI systems that are classified as high-risk according to Article 6 AI Act and thus require human oversight following Article 14 AI Act. For the reasons laid out above, reactive human oversight reaches structural limits with agentic AI systems. Still, Article 14 AI Act requires human oversight to be effective \citep[Art. 14]{euaiact2024}, even for AI agents. Our framework demonstrates how anticipatory oversight of agentic AI systems is compatible with the requirements for effective oversight in Article 14 AI Act. 
The architectural desiderata laid out above (\Cref{subsec:desiderata}) can be mapped onto the enabling measures listed in Article 14(4) AI Act. Inspectability corresponds to Article 14(4)(a) AI Act, i.e., the ability of human oversight personnel \enquote{to properly understand the relevant capacities and limitations of the high-risk AI system and be able to duly monitor its operation, including in view of detecting and addressing anomalies, dysfunctions and unexpected performance}. Articulability facilitates Article 14(4)(c) AI Act, the ability \enquote{to correctly interpret the high-risk AI system's output, taking into account, for example, the interpretation tools and methods available}. Modifiability, together with the substrate's alignment and containment mechanism families (\Cref{subsec:normative-agenda})---particularly external containment, which implements the tripwires, uncertainty-threshold detectors, and triggered hand-offs through which reactive oversight is invoked---promotes Article 14(4)(d) and (e) AI Act, the ability \enquote{to decide, in any particular situation, not to use the high-risk AI system or to otherwise disregard, override or reverse the output of the high-risk AI system} and to \enquote{intervene in the operation of the high-risk AI system or interrupt the system through a `stop' button or a similar procedure that allows the system to come to a halt in a safe state}. The mapping of our architectural desiderata to the enabling measures listed in Article 14(4) AI Act specifies how our framework for anticipatory oversight is compatible with the design requirements of human oversight in the AI Act.

While the anticipatory framework is prima facie compatible with the substantive design requirements for human oversight in Article 14 AI Act, it raises the question of who shall be responsible for anticipatory oversight. Our default assumption throughout the paper has been that anticipatory oversight is a practice of deployers as principals of the agentic system. Article 14 AI Act establishes obligations for providers of AI systems. This is only seemingly a mismatch. First, while Article 14 AI Act obliges providers (those entities who develop an AI system and place it on the market or put it into service, Art. 3(3) AI Act), Article 26(2) AI Act requires deployers (those entities who use an AI system, Art. 3(4) AI Act) to assign human oversight to natural persons who have the necessary skills and competence. The AI Act thus divides the responsibility for enabling effective human oversight between the provider and the deployer, with the former being responsible for development and design and the latter for organizational implementation \citep{laux2025automation}. Furthermore, Article 26(1) AI Act obliges the deployer to adopt appropriate technical and organizational measures to ensure that the AI system is used in accordance with the instructions for use accompanying the systems. Articulating the agenda for anticipatory oversight in the specification–runtime–inspection spiral outlined above could thus fall under Article 26 AI Act and be an obligation for the deployer. However, the effectiveness requirement of Article 14 AI Act would still establish corresponding design obligations for providers of agentic AI systems---if we assume that reactive oversight alone is not effective enough and requires complementary anticipatory oversight. 

Second, Article 25 anticipates the dynamic nature of responsibilities within the AI value chain: a deployer can become a provider under one of three alternative circumstances, including a \enquote{substantial modification} of a high-risk system \cite{finck2026eu}. Given the current legal uncertainty about how that threshold should be interpreted \cite{finck2026eu}, it remains an open question whether a deployer's articulation of the agenda for anticipatory oversight would itself trigger Article 25.

Third, the framework is agnostic about whether anticipatory oversight is exclusively a deployer-side practice or whether providers carry partial anticipatory responsibilities through default agendas, escalation mechanisms, or containment shields shipped with the system. What the framework insists on is that anticipatory oversight is not \emph{architectural goodwill}---something the deployer may voluntarily take up because she chooses to---but a specifiable practice that requires a layered regime of governance: provider obligations under Articles 8--15 condition the architectural possibility of anticipatory oversight by constraining the alignment and containment mechanisms; deployer obligations under Article 26 condition the practice by constraining how the principal uses the system; post-market obligations under Articles 20, 26(5), and 72 condition the loop between deployment and provider learning. Anticipatory oversight is one layer of this layered regime, not a stand-alone solution.

\medskip
\emph{Architectural research.} The three desiderata of \Cref{subsec:desiderata} formulate a research agenda for the three mechanism families set out in \Cref{subsec:normative-agenda}---alignment, internal containment, and external containment. The relevant question for any given mechanism is not whether it is \enquote{aligned} in an undifferentiated sense but whether it supports Articulability, Inspectability, and Modifiability in the substantive readings we have developed. On the alignment side, the question becomes how foundation-model alignment techniques can deliver dispositions that track an articulated agenda---a structure of normative reasons rather than a flat rule-set---and remain modifiable as the agenda is refined. On the internal-containment side, neuro-symbolic architectures that decouple normative reasoning from instrumental decision-making provide one direction in which this question is being explored: \citet{jahn2026grace}, for instance, develop an neuro-symbolic containment architecture in which a Moral Module formulates permissible action types through reason-based deontic logic, a Decision-Making Module handles instrumental optimisation within those bounds, and a Guard enforces compliance at runtime---an architecture which instantiates a reason-based, defeasible, and revisable specification of the kind the desiderata require.  
On the external-containment side, the question concerns the design of runtime safeguards: tripwires, uncertainty-threshold detectors, triggered hand-off protocols, and a new kind of institutional and organisational infrastructures that operate at the deployment perimeter. Which family, or which combination, will turn out the more tractable route for any given deployment is an open question; what the framework contributes is a specification of what such routes must deliver to count as anticipatory oversight at all.

\medskip
\emph{Empirical research.} Several empirical questions follow directly from the framework. How, in practice, do principals articulate agendas? Which articulation interfaces support the structure of normative reasons developed in \Cref{subsec:reasons} and which collapse them into flat rule lists? How does inspection scale with trajectory volume, and at what point does the oversight personnel lose vigilance? Which institutional configurations underwrite stakeholder contestability and which preempt it? 

Two recent strands of empirical work provide both methodological starting points and concrete diagnostic tools: red-teaming studies of deployed agentic systems \citep{shapira2026agents} make compliance failures visible at scale, and multi-dimensional assessments of moral competence in language models \citep{kilov2026discerning} identify where agent-side reasons-tracking fails. The literature on reactive oversight effectiveness \citep{sterz2024quest,langer2025signal} and supervisory control \cite{sheridan2021supervisory}  
provide further theoretical and methodological grounding as well as inspiration regarding how to empirically approach anticipatory human oversight effectiveness.

\section{Conclusion}
\label{sec:conclusion}

For agentic AI, reactive oversight reaches structural limits: dense over-the-shoulder oversight defeats the autonomy that motivates the deployment, while loose ex-post oversight leaves cumulative-trajectory failures unaddressed. The anticipatory mode developed in this paper is the operationalisation of distal-reason tracking for agentic AI: the principal specifies, before the agent's engagement, a normative agenda that structures the conditions under which the agent acts, and refines it iteratively through inspection. The framework reconstructs both conditions of Meaningful Human Control for the agentic case---tracking via the agenda (\Cref{sec:agenda}) and tracing via the responsibility architecture the framework makes available by design (\Cref{sec:responsibility}).

The anticipatory mode is itself the discharge of a role-grounded prospective obligation; backward-looking responsibility takes the form of strict moral answerability---rationalistic, relational, and holding regardless of fault in virtue of the principal's prior opportunity for precaution. The agenda's substrate, across alignment, internal containment, and external containment, must satisfy three architectural desiderata: articulability with the structure of normative reasons, inspectability of trajectories against the agenda, and modifiability at the granularity of considerations. 
Bridging failure modes (\Cref{sec:failures}) connect this diagnosis to the responsibility constellations, locating where failure arises and who must answer; their mapping onto regulatory loci under the EU AI Act we sketch but leave to dedicated treatment. Limits and preconditions (\Cref{sec:limits}) acknowledge what the framework presupposes but does not itself secure.

The framework is a \emph{normative} specification; satisfiability is an empirical question. The record on deployed agentic systems suggests the desiderata are not currently met in their non-trivial readings. The implication is direct: if anticipatory oversight is necessary for effective oversight of high-risk agentic systems, and the desiderata cannot be satisfied in present practice, the appropriate response is to defer such deployments until they can. The architectural and empirical research agenda of \Cref{sec:implications} marks the route by which that deferral might be lifted.

\section*{Declarations}

\noindent\emph{Funding.} The work of Kevin Baum and Markus Langer was partially funded by the Deutsche Forschungsgemeinschaft (DFG, German Research Foundation) -- Project-ID 389792660 -- TRR~248 (\url{https://perspicuous-computing.science}). Kevin Baum's work was further supported by the German Federal Ministry of Education and Research (BMBF) and its successor, the Federal Ministry of Research, Technology and Space (BMFTR), within the projects MAC-MERLin (grant no.~16IW24007) and SAgA (grant no.~16IW26005), and by the European Regional Development Fund (ERDF) and the Saarland within the project ToCERTAIN (ID: EFRE-AuF-0000942).

\noindent\emph{Competing interests.} The author(s) declare no competing interests.

\noindent\emph{Ethics approval.} Not applicable.

\noindent\emph{Data availability.} Not applicable.

\noindent\emph{Author contributions.} 
K.B. is first and senior author. He conceived the project, developed the framework, and wrote the manuscript. K.B. and M.K. jointly developed the account of responsibility as answerability and its connection to the oversight framework (Section~4). K.B. and M.L. jointly developed the parts drawing on supervisory control, management theory, and principal-agent theory (Section~2.1), with M.L. taking the lead. J.L. was responsible for the regulatory analysis under the EU AI Act (Sections~3.5 and~6). All authors revised the manuscript critically and approved the final version.

\bibliography{references}

@inproceedings{sterz2024quest,
  author    = {Sterz, Sarah and Baum, Kevin and Biewer, Sebastian and Hermanns, Holger and Lauber-R{\"o}nsberg, Anne and Meinel, Philip and Langer, Markus},
  title     = {On the Quest for Effectiveness in Human Oversight: Interdisciplinary Perspectives},
  booktitle = {The 2024 ACM Conference on Fairness, Accountability, and Transparency (FAccT '24)},
  pages     = {2495--2507},
  year      = {2024},
  publisher = {ACM},
  address   = {New York, NY, USA},
  doi       = {10.1145/3630106.3659051}
}

@article{langer2025signal,
  author    = {Langer, Markus and Baum, Kevin and Schlicker, Nadine},
  title     = {Effective Human Oversight of {AI}-Based Systems: A Signal Detection Perspective on the Detection of Inaccurate and Unfair Outputs},
  journal   = {Minds and Machines},
  volume    = {35},
  number    = {1},
  pages     = {1},
  year      = {2025},
  doi       = {10.1007/s11023-024-09701-0},
  note      = {Article 1}
}

@unpublished{baum2026causal,
  author    = {Baum, Kevin and Laux, Johann},
  title     = {Constitutive vs.\ Corrective: A Causal Taxonomy of Human Runtime Involvement in {AI} Systems},
  note      = {SSRN preprint, available at \url{https://ssrn.com/abstract=6445598}},
  year      = {2026}
}

@unpublished{gaube2026keeping,
  author    = {Gaube, Susanne and Langer, Markus and Miller, Tim and Baum, Kevin and Dachselt, Raimund and Feit, Anna Maria and Gadiraju, Ujwal and Kaur, Harmanpreet and Keane, Mark T. and Landers, Richard and Laux, Johann and Liao, Q. Vera and Lim, Brian and Onnasch, Linda and Schrills, Tim and Sonenberg, Liz and Tan, Chenhao and Tintarev, Nava and Xiao, Ziang and Zhang, Hanwei},
  title     = {Keeping an Eye on {AI}: A Framework for Effective Human Oversight of {AI} Systems},
  note      = {Preprint, March 2026. Available at \url{https://ssrn.com/abstract=6576159}},
  year      = {2026}
}

@inproceedings{baum2026principles,
  author    = {Baum, Kevin and Uth, Richard and Hermanns, Holger and Kerstan, Sophie and Langer, Markus and Lauber-R{\"o}nsberg, Anne and Meinel, Philip and Stenzel, Laura and Sterz, Sarah and Zhang, Hanwei},
  title     = {The Principal's Principles: Actionable (Personalized) {AI} Alignment as Underexplored {XAI} Application Context},
  booktitle = {Contextualizing Explanations: Proceedings of the 3rd TRR 318 Conference (ConTeX25)},
  publisher = {Bielefeld University Press},
  address   = {Bielefeld},
  year      = {2026},
  doi       = {10.64136/jfoo1236},
  note      = {Extended abstract}
}

@article{baum2022responsibility,
  author    = {Baum, Kevin and Mantel, Susanne and Schmidt, Eva and Speith, Timo},
  title     = {From Responsibility to Reason-Giving Explainable Artificial Intelligence},
  journal   = {Philosophy \& Technology},
  volume    = {35},
  number    = {1},
  pages     = {12},
  year      = {2022},
  doi       = {10.1007/s13347-022-00510-w},
  note      = {Article 12}
}

@inproceedings{faas2026design,
  author    = {Faas, Cedric and Kerstan, Sophie and Uth, Richard and Langer, Markus and Feit, Anna Maria},
  title     = {Design Considerations for Human Oversight of {AI}: Insights from Co-Design Workshops and Work Design Theory},
  booktitle = {Proceedings of the 31st International Conference on Intelligent User Interfaces (IUI '26)},
  year      = {2026},
  publisher = {ACM},
  address   = {New York, NY, USA},
  doi       = {10.1145/3742413.3789100}
}

@inproceedings{jahn2026grace,
  author    = {Jahn, Felix and Muskalla, Yannic and Dargasz, Lisa and Schramowski, Patrick and Baum, Kevin},
  title     = {Breaking Up with Normatively Monolithic Agency with {GRACE}: A Reason-Based Neuro-Symbolic Architecture for Safe and Ethical {AI} Alignment},
  booktitle = {Proceedings of the 2nd Annual Conference of the International Association for Safe \& Ethical {AI} (IASEAI '26)},
  year      = {2026},
  doi       = {10.48550/arXiv.2601.10520},
  note      = {arXiv:2601.10520}
}

@article{santoni2018meaningful,
  author    = {{Santoni de Sio}, Filippo and {van den Hoven}, Jeroen},
  title     = {Meaningful Human Control over Autonomous Systems: A Philosophical Account},
  journal   = {Frontiers in Robotics and AI},
  volume    = {5},
  pages     = {15},
  year      = {2018},
  doi       = {10.3389/frobt.2018.00015}
}

@article{mecacci2020meaningful,
  author    = {Mecacci, Giulio and {Santoni de Sio}, Filippo},
  title     = {Meaningful Human Control as Reason-Responsiveness: The Case of Dual-Mode Vehicles},
  journal   = {Ethics and Information Technology},
  volume    = {22},
  number    = {2},
  pages     = {103--115},
  year      = {2020},
  doi       = {10.1007/s10676-019-09519-w}
}

@article{santoni2021four,
  author    = {{Santoni de Sio}, Filippo and Mecacci, Giulio},
  title     = {Four Responsibility Gaps with Artificial Intelligence: Why They Matter and How to Address Them},
  journal   = {Philosophy \& Technology},
  volume    = {34},
  number    = {4},
  pages     = {1057--1084},
  year      = {2021},
  doi       = {10.1007/s13347-021-00450-x}
}

@article{green2022flaws,
  author    = {Green, Ben},
  title     = {The Flaws of Policies Requiring Human Oversight of Government Algorithms},
  journal   = {Computer Law \& Security Review},
  volume    = {45},
  pages     = {105681},
  year      = {2022},
  doi       = {10.1016/j.clsr.2022.105681}
}

@article{elish2019moral,
  author    = {Elish, Madeleine Clare},
  title     = {Moral Crumple Zones: Cautionary Tales in Human-Robot Interaction},
  journal   = {Engaging Science, Technology, and Society},
  volume    = {5},
  pages     = {40--60},
  year      = {2019},
  doi       = {10.17351/ests2019.260}
}

@article{laux2023institutionalised,
  author    = {Laux, Johann},
  title     = {Institutionalised Distrust and Human Oversight of Artificial Intelligence: Toward a Democratic Design of {AI} Governance under the {European Union AI Act}},
  journal   = {AI \& Society},
  volume    = {39},
  number    = {6},
  pages     = {2853--2866},
  year      = {2024},
  doi       = {10.1007/s00146-023-01794-y}
}

@article{botti2025agentic,
  title={Agentic AI and Multiagentic: Are We Reinventing the Wheel?},
  eprint={2506.01463},
  archivePrefix={arXiv},
  year={2025},
  author = {Vicent Botti},
  primaryClass={cs.MA},
  doi= {10.48550/arXiv.2506.01463} 
}

@article{laux2025automation,
  author    = {Laux, Johann and Ruschemeier, Hannah},
  title     = {Automation Bias in the {AI Act}: On the Legal Implications of Attempting to De-Bias Human Oversight of {AI}},
  journal   = {European Journal of Risk Regulation},
  volume    = {16},
  number    = {4},
  pages     = {1519--1534},
  year      = {2025},
  doi       = {10.1017/err.2025.10033}
}

@unpublished{fink2025oversight,
  author    = {Fink, Melanie},
  title     = {Human Oversight under {Article 14} of the {EU AI Act}},
  note      = {SSRN preprint. Forthcoming in Malgieri et al.\ (eds.), {AI Act Commentary}, Hart-Bloomsbury},
  year      = {2025}
}

@article{enarsson2023oversight,
  author    = {Enarsson, Therese and Enqvist, Lena and Naarttij{\"a}rvi, Markus},
  title     = {`Human Oversight' in the {EU} Artificial Intelligence Act: What, When and by Whom?},
  journal   = {Law, Innovation and Technology},
  volume    = {15},
  number    = {2},
  pages     = {508--535},
  year      = {2023},
  doi       = {10.1080/17579961.2023.2245683}
}

@unpublished{passi2025agentic,
  author    = {Passi, Samir},
  title     = {Agentic {AI} Has a Human Oversight Problem},
  note      = {SSRN preprint},
  year      = {2025},
  doi       = {10.2139/ssrn.5529058}
}

@article{gabriel2024ethics,
  author    = {Gabriel, Iason and Manzini, Arianna and Keeling, Geoff and Hendricks, Lisa Anne and Rieser, Verena and Iqbal, Hasan and Toma{\v{s}}ev, Nenad and Ktena, Ira and Kenton, Zachary and Rodriguez, Mikel},
  title     = {The Ethics of Advanced {AI} Assistants},
  journal   = {arXiv preprint},
  year      = {2024},
  note      = {arXiv:2404.16244}
}

@inproceedings{chan2024visibility,
  author    = {Alan Chan and Carson Ezell and Max Kaufmann and Kevin Wei and Lewis Hammond and Herbie Bradley and Emma Bluemke and Nitarshan Rajkumar and David Krueger and Noam Kolt and Lennart Heim and Markus Anderljung},
  title     = {Visibility into {AI} Agents},
  booktitle = {Proceedings of the 2024 ACM Conference on Fairness, Accountability, and Transparency (FAccT '24)},
  year      = {2024},
  publisher = {ACM},
  address   = {New York, NY, USA},
  doi       = {10.1145/3630106.3658948}
}

@article{kasirzadeh2025characterizing,
  author    = {Kasirzadeh, Atoosa and Gabriel, Iason},
  title     = {Characterizing {AI} Agents for Alignment and Governance},
  journal   = {arXiv preprint},
  year      = {2025},
  note      = {arXiv:2504.21848}
}

@article{mitchell2025fully,
  author    = {Mitchell, Margaret and Ghosh, Avijit and Luccioni, Alexandra Sasha and Pistilli, Giada},
  title     = {Fully Autonomous {AI} Agents Should Not Be Developed},
  journal   = {arXiv preprint},
  year      = {2025},
  note      = {arXiv:2502.02649}
}

@article{kiener2024strict,
  author    = {Kiener, Maximilian},
  title     = {Strict Moral Answerability},
  journal   = {Ethics},
  volume    = {134},
  number    = {3},
  pages     = {360--386},
  year      = {2024},
  doi       = {10.1086/728635}
}

@article{kiener2025abundance,
  author    = {Kiener, Maximilian},
  title     = {{AI} and Responsibility: No Gap, but Abundance},
  journal   = {Journal of Applied Philosophy},
  volume    = {42},
  number    = {1},
  pages     = {357--374},
  year      = {2025},
  doi       = {10.1111/japp.12765}
}

@incollection{vandepoel2011relation,
  author    = {{Van de Poel}, Ibo},
  title     = {The Relation between Forward-Looking and Backward-Looking Responsibility},
  booktitle = {Moral Responsibility: Beyond Free Will and Determinism},
  editor    = {Vincent, Nicole A. and {Van de Poel}, Ibo and {Van den Hoven}, Jeroen},
  publisher = {Springer},
  address   = {Dordrecht},
  pages     = {37--52},
  year      = {2011},
  doi       = {10.1007/978-94-007-1878-4_3}
}

@incollection{williams1981moral,
  author    = {Williams, Bernard},
  title     = {Moral Luck},
  booktitle = {Moral Luck: Philosophical Papers, 1973--1980},
  publisher = {Cambridge University Press},
  address   = {Cambridge},
  pages     = {20--39},
  year      = {1981}
}

@incollection{nagel1979moral,
  author    = {Nagel, Thomas},
  title     = {Moral Luck},
  booktitle = {Mortal Questions},
  publisher = {Cambridge University Press},
  address   = {Cambridge},
  pages     = {24--38},
  year      = {1979}
}

@book{honore1999responsibility,
  author    = {Honor{\'e}, Tony},
  title     = {Responsibility and Fault},
  publisher = {Hart Publishing},
  address   = {Oxford},
  year      = {1999}
}

@book{applbaum1999ethics,
  author    = {Applbaum, Arthur Isak},
  title     = {Ethics for Adversaries: The Morality of Roles in Public and Professional Life},
  publisher = {Princeton University Press},
  address   = {Princeton, NJ},
  year      = {1999}
}

@article{bainbridge1983ironies,
  author    = {Bainbridge, Lisanne},
  title     = {Ironies of Automation},
  journal   = {Automatica},
  volume    = {19},
  number    = {6},
  pages     = {775--779},
  year      = {1983},
  doi       = {10.1016/0005-1098(83)90046-8}
}

@book{sheridan1992telerobotics,
  author    = {Sheridan, Thomas B.},
  title     = {Telerobotics, Automation, and Human Supervisory Control},
  publisher = {MIT Press},
  address   = {Cambridge, MA},
  year      = {1992}
}

@incollection{sheridan2021supervisory,
  author    = {Sheridan, Thomas B.},
  title     = {Human Supervisory Control of Automation},
  booktitle = {Handbook of Human Factors and Ergonomics},
  editor    = {Salvendy, Gavriel and Karwowski, Waldemar},
  edition   = {5th},
  publisher = {Wiley},
  address   = {Hoboken, NJ},
  pages     = {736--760},
  year      = {2021}
}

@techreport{sheridan1978human,
  author    = {Sheridan, Thomas B. and Verplank, William L.},
  title     = {Human and Computer Control of Undersea Teleoperators},
  institution = {Man-Machine Systems Laboratory, MIT},
  address   = {Cambridge, MA},
  year      = {1978}
}

@article{parasuraman2010complacency,
  author    = {Parasuraman, Raja and Manzey, Dietrich H.},
  title     = {Complacency and Bias in Human Use of Automation: An Attentional Integration},
  journal   = {Human Factors},
  volume    = {52},
  number    = {3},
  pages     = {381--410},
  year      = {2010},
  doi       = {10.1177/0018720810376055}
}

@article{gabriel2020values,
  author    = {Gabriel, Iason},
  title     = {Artificial Intelligence, Values, and Alignment},
  journal   = {Minds and Machines},
  volume    = {30},
  number    = {3},
  pages     = {411--437},
  year      = {2020},
  doi       = {10.1007/s11023-020-09539-2}
}

@article{gabriel2025principle,
  author    = {Gabriel, Iason and Keeling, Geoff},
  title     = {A Matter of Principle? {AI} Alignment as the Fair Treatment of Claims},
  journal   = {Philosophical Studies},
  volume    = {182},
  number    = {7},
  pages     = {1951--1973},
  year      = {2025},
  doi       = {10.1007/s11098-025-02300-4}
}

@book{horty2012reasons,
  author    = {Horty, John F.},
  title     = {Reasons as Defaults},
  publisher = {Oxford University Press},
  address   = {Oxford},
  year      = {2012}
}

@book{drucker1954practice,
  author    = {Drucker, Peter F.},
  title     = {The Practice of Management},
  publisher = {Harper \& Row},
  address   = {New York},
  year      = {1954}
}

@article{jensen1976,
  author    = {Jensen, Michael C. and Meckling, William H.},
  title     = {Theory of the Firm: Managerial Behavior, Agency Costs and Ownership Structure},
  journal   = {Journal of Financial Economics},
  volume    = {3},
  number    = {4},
  pages     = {305--360},
  year      = {1976},
  doi       = {10.1016/0304-405X(76)90026-X}
}

@article{eisenhardt1989agency,
  author    = {Eisenhardt, Kathleen M.},
  title     = {Agency Theory: An Assessment and Review},
  journal   = {Academy of Management Review},
  volume    = {14},
  number    = {1},
  pages     = {57--74},
  year      = {1989},
  doi       = {10.5465/amr.1989.4279003}
}

@book{bratman1987intention,
  author    = {Bratman, Michael E.},
  title     = {Intention, Plans, and Practical Reason},
  publisher = {Harvard University Press},
  address   = {Cambridge, MA},
  year      = {1987}
}

@book{anscombe1957intention,
  author    = {Anscombe, G. E. M.},
  title     = {Intention},
  publisher = {Basil Blackwell},
  address   = {Oxford},
  year      = {1957}
}

@InProceedings{10.1007/978-3-032-01377-4_8,
author="Baum, Kevin",
editor="Steffen, Bernhard",
title={{Disentangling AI Alignment: A Structured Taxonomy Beyond Safety and Ethics}},
booktitle="Bridging the Gap Between AI and Reality",
year="2026",
publisher="Springer Nature Switzerland",
address="Cham",
pages="158--173",
isbn="978-3-032-01377-4"
}

@book{davidson1980essays,
  author    = {Davidson, Donald},
  title     = {Essays on Actions and Events},
  publisher = {Clarendon Press},
  address   = {Oxford},
  year      = {1980}
}

@book{mele1992springs,
  author    = {Mele, Alfred R.},
  title     = {Springs of Action: Understanding Intentional Behavior},
  publisher = {Oxford University Press},
  address   = {New York},
  year      = {1992}
}

@misc{euaiact2024,
  author       = {{European Parliament and Council of the European Union}},
  title        = {{Regulation (EU) 2024/1689 of the European Parliament and of the Council of 13 June 2024 laying down harmonised rules on artificial intelligence (Artificial Intelligence Act)}},
  year         = {2024},
  howpublished = {Official Journal of the European Union, L series, 12 July 2024},
  url          = {https://eur-lex.europa.eu/eli/reg/2024/1689/oj/eng}
}

@book{raz1990practical,
  author    = {Raz, Joseph},
  title     = {Practical Reason and Norms},
  edition   = {2nd},
  publisher = {Princeton University Press},
  address   = {Princeton, NJ},
  year      = {1990},
  note      = {First edition published 1975 by Hutchinson, London}
}

@book{dancy2000practical,
  author    = {Dancy, Jonathan},
  title     = {Practical Reality},
  publisher = {Oxford University Press},
  address   = {Oxford},
  year      = {2000}
}

@book{dancy2004ethics,
  author    = {Dancy, Jonathan},
  title     = {Ethics Without Principles},
  publisher = {Clarendon Press},
  address   = {Oxford},
  year      = {2004}
}

@book{rawls1993political,
  author    = {Rawls, John},
  title     = {Political Liberalism},
  publisher = {Columbia University Press},
  address   = {New York},
  year      = {1993}
}

@book{macaskill2020uncertainty,
  author    = {MacAskill, William and Bykvist, Krister and Ord, Toby},
  title     = {Moral Uncertainty},
  publisher = {Oxford University Press},
  address   = {Oxford},
  year      = {2020}
}

@book{lockhart2000moral,
  author    = {Lockhart, Ted},
  title     = {Moral Uncertainty and Its Consequences},
  publisher = {Oxford University Press},
  address   = {New York},
  year      = {2000}
}

@article{coeckelbergh2020,
  author  = {Coeckelbergh, Mark},
  title   = {Artificial Intelligence, Responsibility Attribution, and a Relational Justification of Explainability},
  journal = {Science and Engineering Ethics},
  volume  = {26},
  number  = {4},
  pages   = {2051--2068},
  year    = {2020},
  doi     = {10.1007/s11948-019-00146-8}
}

@book{duff2007,
  author    = {Duff, Antony},
  title     = {Answering for Crime: Responsibility and Liability in the Criminal Law},
  publisher = {Hart},
  address   = {Oxford},
  year      = {2007}
}

@book{fischer1998,
  author    = {Fischer, John Martin and Ravizza, Mark},
  title     = {Responsibility and Control: A Theory of Moral Responsibility},
  publisher = {Cambridge University Press},
  address   = {Cambridge},
  year      = {1998}
}

@article{hardimon1994,
  author  = {Hardimon, Michael O.},
  title   = {Role Obligations},
  journal = {Journal of Philosophy},
  volume  = {91},
  number  = {7},
  pages   = {333--363},
  year    = {1994}
}

@article{hohfeld1913,
  author  = {Hohfeld, Wesley Newcomb},
  title   = {Some Fundamental Legal Conceptions as Applied in Judicial Reasoning},
  journal = {Yale Law Journal},
  volume  = {23},
  number  = {1},
  pages   = {16--59},
  year    = {1913}
}

@article{kiener2022,
  author  = {Kiener, Maximilian},
  title   = {Can We Bridge AI's Responsibility Gap at Will?},
  journal = {Ethical Theory and Moral Practice},
  volume  = {25},
  number  = {4},
  pages   = {575--593},
  year    = {2022},
  doi     = {10.1007/s10677-022-10313-9}
}

@article{koenigs2022,
  author  = {K{\"o}nigs, Peter},
  title   = {Artificial Intelligence and Responsibility Gaps: What Is the Problem?},
  journal = {Ethics and Information Technology},
  volume  = {24},
  number  = {3},
  pages   = {1--11},
  year    = {2022},
  doi     = {10.1007/s10676-022-09643-0}
}

@article{matthias2004,
  author  = {Matthias, Andreas},
  title   = {The Responsibility Gap: Ascribing Responsibility for the Actions of Learning Automata},
  journal = {Ethics and Information Technology},
  volume  = {6},
  number  = {3},
  pages   = {175--183},
  year    = {2004},
  doi     = {10.1007/s10676-004-3422-1}
}

@book{nyholm2020,
  author    = {Nyholm, Sven},
  title     = {Humans and Robots: Ethics, Agency, and Anthropomorphism},
  publisher = {Rowman \& Littlefield},
  address   = {Lanham},
  year      = {2020}
}

@book{robichaud2017,
  editor    = {Robichaud, Philip and Wieland, Jan Willem},
  title     = {Responsibility: The Epistemic Condition},
  publisher = {Oxford University Press},
  address   = {Oxford},
  year      = {2017}
}

@book{scanlon1998,
  author    = {Scanlon, T. M.},
  title     = {What We Owe to Each Other},
  publisher = {Belknap Press of Harvard University Press},
  address   = {Cambridge, MA},
  year      = {1998}
}

@incollection{shoemaker2013,
  author    = {Shoemaker, David},
  title     = {On Criminal and Moral Responsibility},
  booktitle = {Oxford Studies in Normative Ethics, Volume 3},
  editor    = {Timmons, Mark},
  publisher = {Oxford University Press},
  address   = {Oxford},
  pages     = {154--178},
  year      = {2013}
}

@book{shoemaker2015,
  author    = {Shoemaker, David},
  title     = {Responsibility from the Margins},
  publisher = {Oxford University Press},
  address   = {Oxford},
  year      = {2015}
}

@article{smith2015,
  author  = {Smith, Angela},
  title   = {Responsibility as Answerability},
  journal = {Inquiry},
  volume  = {58},
  number  = {2},
  pages   = {99--126},
  year    = {2015},
  doi     = {10.1080/0020174X.2015.986851}
}

@article{sparrow2007,
  author  = {Sparrow, Robert},
  title   = {Killer Robots},
  journal = {Journal of Applied Philosophy},
  volume  = {24},
  number  = {1},
  pages   = {62--77},
  year    = {2007},
  doi     = {10.1111/j.1468-5930.2007.00346.x}
}

@book{strawson1962,
  author    = {Strawson, P. F.},
  title     = {Freedom and Resentment},
  publisher = {British Academy},
  address   = {London},
  year      = {1962}
}

@book{watson2004,
  author    = {Watson, Gary},
  title     = {Agency and Answerability: Selected Essays},
  publisher = {Oxford University Press},
  address   = {Oxford},
  year      = {2004}
}

@book{williams1981,
  author    = {Williams, Bernard},
  title     = {Moral Luck: Philosophical Papers, 1973--1980},
  publisher = {Cambridge University Press},
  address   = {Cambridge},
  year      = {1981}
}

@article{zimmerman1985,
  author  = {Zimmerman, Michael J.},
  title   = {Sharing Responsibility},
  journal = {American Philosophical Quarterly},
  volume  = {22},
  number  = {2},
  pages   = {115--122},
  year    = {1985}
}

@book{Nozick1981,
  title = {Philosophical Explanations},
  author = {Nozick, Robert},
  year = {1981},
  publisher = {Harvard University Press},
  address = {Cambridge, MA},
  isbn = {9780674664791}
}

@book{finck2026eu,
  author = {Finck, Michèle},
  title = {The {EU} Artificial Intelligence Act: {A} Commentary},
  publisher = {Oxford University Press},
  address = {Oxford},
  year = {2026},
  isbn = {9780198925729}
}

@article{Dung2025-DUNUAA,
	author = {Leonard Dung},
	doi = {10.1093/pq/pqae010},
	journal = {Philosophical Quarterly},
	number = {2},
	pages = {450--472},
	title = {{Understanding Artificial Agency}},
	volume = {75},
	year = {2025}
}

@article{shapira2026agents,
  author = {Shapira, Natalie and Wendler, Chris and Yen, Avery and Sarti, Gabriele and Pal, Koyena and Floody, Olivia and Belfki, Adam and Loftus, Alex and Jannali, Aditya Ratan and Prakash, Nikhil and Cui, Jasmine and Rogers, Giordano and Brinkmann, Jannik and Rager, Can and Zur, Amir and Ripa, Michael and Sankaranarayanan, Aruna and Atkinson, David and Gandikota, Rohit and Fiotto-Kaufman, Jaden and Hwang, EunJeong and Orgad, Hadas and Sahil, P Sam and Taglicht, Negev and Shabtay, Tomer and Ambus, Atai and Alon, Nitay and Oron, Shiri and Gordon-Tapiero, Ayelet and Kaplan, Yotam and Shwartz, Vered and Rott Shaham, Tamar and Riedl, Christoph and Mirsky, Reuth and Sap, Maarten and Manheim, David and Ullman, Tomer and Bau, David},
  title = {Agents of {C}haos},
  journal = {arXiv},
  year = {2026},
  eprint = {2602.20021},
  archivePrefix = {arXiv},
  primaryClass = {cs.AI},
  month = feb,
  url = {https://arxiv.org/abs/2602.20021},
  note = {Preprint}
}

@article{kozlovski2025reasons,
  author = {Kozlovski, A.},
  title = {{R}easons underdetermination in meaningful human control},
  journal = {Ethics and Information Technology},
  volume = {27},
  article = {59},
  year = {2025},
  doi = {10.1007/s10676-025-09858-x},
  url = {https://doi.org/10.1007/s10676-025-09858-x}
}

@misc{li2026controllability,
      title={{Position: AI Safety Requires Effective Controllability}}, 
      author={Yige Li and Yunhao Feng and Jun Sun},
      year={2026},
      eprint={2605.27117},
      archivePrefix={arXiv},
      primaryClass={cs.AI},
      url={https://arxiv.org/abs/2605.27117}, 
}

@article{hindriks2023control,
  author = {Hindriks, F. and Veluwenkamp, H.},
  title = {The risks of autonomous machines: from responsibility gaps to control gaps},
  journal = {Synthese},
  volume = {201},
  article = {21},
  year = {2023},
  doi = {10.1007/s11229-022-04001-5},
  url = {https://doi.org/10.1007/s11229-022-04001-5}
}

@article{singh2025architecting,
  author = {Singh, A. and Szajnfarber, Z.},
  title = {Architecting {H}uman-{AI} Systems for Effective Collaboration and Oversight: {M}aking Sense of {H}uman/{AI}-in/on/over/{U}nder/{A}long-the-{L}oop},
  journal = {Systems Engineering},
  volume = {29},
  article = {e70024},
  year = {2025},
  doi = {10.1002/sys.70024},
  url = {https://doi.org/10.1002/sys.70024}
}

@article{veluwenkamp2022reasons,
  author = {Veluwenkamp, H.},
  title = {{R}easons for {M}eaningful {H}uman {C}ontrol},
  journal = {Ethics and Information Technology},
  volume = {24},
  article = {51},
  year = {2022},
  doi = {10.1007/s10676-022-09673-8},
  url = {https://doi.org/10.1007/s10676-022-09673-8}
}

@article{kilov2026discerning,
  author = {Kilov, Daniel and Hendy, Caroline and Yanik Guyot, Secil and Snoswell, Aaron J. and Lazar, Seth},
  title = {{D}iscerning {W}hat {M}atters: {A} {M}ulti-{D}imensional {A}ssessment of {{M}oral {C}ompetence in {LLM}s}},
  journal = {arXiv},
  year = {2025},
  month = jun,
  eprint = {2506.13082},
  archivePrefix = {arXiv},
  primaryClass = {cs.AI},
  url = {https://arxiv.org/abs/2506.13082},
  note = {Preprint}
}

@article{lazaros2026human,
  author = {Lazaros, Konstantinos and Vrahatis, Aristidis G. and Kotsiantis, Sotiris},
  title = {{H}uman-in-the-{L}oop {A}rtificial {I}ntelligence: {A} {S}ystematic {R}eview of {C}oncepts, {M}ethods, and {A}pplications},
  journal = {Entropy},
  volume = {28},
  number = {4},
  pages = {377},
  year = {2026},
  doi = {10.3390/e28040377},
  url = {https://doi.org/10.3390/e28040377}
}

\end{document}